%% file: main.tex
\documentclass{Nature}
\usepackage[normalem]{ulem}
\usepackage{graphicx}
\usepackage{amssymb}
\usepackage{amsmath}
\usepackage{graphicx}
\usepackage{dcolumn}
\usepackage{bm}
\usepackage{hyperref}
\usepackage{siunitx}
\usepackage{graphicx}
\usepackage{xcolor}
\usepackage{xspace}
\usepackage[capitalize]{cleveref}
\usepackage{multirow}

\makeatletter
\let\saved@includegraphics\includegraphics
\AtBeginDocument{\let\includegraphics\saved@includegraphics}

\makeatletter

\makeatletter

\newcommand{\changedtwo}[1]{{\color{black} #1}}

\newcommand{\params}{\ensuremath{\boldsymbol{\theta}}}
\newcommand{\model}{\ensuremath{\Omega}}

\newcommand{\order}[1]{\ensuremath{\mathcal{O}(#1)}}

\newcommand{\data}{\ensuremath{\mathbf{d}}\xspace}
\newcommand{\likelihood}{\ensuremath{\mathcal{L}}\xspace}
\newcommand{\prior}{\ensuremath{\pi}\xspace}
\newcommand{\evidence}{\ensuremath{\mathcal{Z}}\xspace}
\newcommand{\bayesfactor}{\ensuremath{\rm BF}\xspace}
\newcommand{\odds}{\ensuremath{\mathcal{O}}\xspace}

\newcommand{\lambdatilde}{\ensuremath{\tilde{\Lambda}}\xspace}
\newcommand{\deltalambdatilde}{\ensuremath{\Delta\tilde{\Lambda}}\xspace}

\newcommand{\bilbyMCMC}{\textsc{Bilby-MCMC}\xspace}
\newcommand{\dynesty}{\textsc{dynesty}\xspace}

\newcommand{\bayesline}{\textsc{BayesLine}\xspace}

\newcommand{\imrDNRTnoxspace}{\texttt{IMRPhenomD\_NRTidalv2}}
\newcommand{\imrDNRT}{\imrDNRTnoxspace\xspace}

\newcommand{\seoNRTnoxspace}{\texttt{SEOBNRv4\_ROM\_NRTidalv2}}
\newcommand{\seoNRT}{\seoNRTnoxspace\xspace}

\newcommand{\seoTnoxspace}{\texttt{SEOBNRv4T\_surrogate}}
\newcommand{\seoT}{\seoTnoxspace\xspace}

\newcommand{\teonoxspace}{\texttt{TEOBResumS}}
\newcommand{\teo}{\teonoxspace\xspace}

\newcommand{\seventeennoxspace}{GW170817}
\newcommand{\seventeen}{\seventeennoxspace\xspace}

\newcommand{\nineteennoxspace}{GW190425}
\newcommand{\nineteen}{\nineteennoxspace\xspace}

\newcommand{\subthresnoxspace}{GW200311\_103121}
\newcommand{\subthres}{\subthresnoxspace\xspace}

\makeatother

\usepackage{expl3}[2012-07-08]
\ExplSyntaxOn
\cs_new_eq:NN \fpeval \fp_eval:n
\ExplSyntaxOff

\title{The use of hypermodels to understand binary neutron star collisions}

\author{Gregory Ashton$^{1,2}$ and Tim Dietrich$^{3,4}$}

\begin{document}
\input{macros}

\maketitle

\begin{affiliations}
\item {Department of Physics, Royal Holloway, University of London, TW20 0EX, United Kingdom}
\item {University of Portsmouth, Institute of Cosmology and Gravitation, Portsmouth PO1 3FX, United Kingdom}
\item{Institute for Physics and Astronomy, University of Potsdam, D-14476 Potsdam, Germany}
\item{Max Planck Institute for Gravitational Physics (Albert Einstein Institute), Am M{\"u}hlenberg 1, D-14476 Potsdam, Germany}
\end{affiliations}

\begin{abstract}
Gravitational waves from the collision of binary neutron stars provide a unique opportunity to study the behaviour of supranuclear matter, the fundamental properties of gravity, and the cosmic history of our Universe. However, given the complexity of Einstein's Field Equations, theoretical models that enable source-property inference suffer from systematic uncertainties due to simplifying assumptions. We develop a hypermodel approach to compare and measure the uncertainty \changedtwo{of} gravitational-wave approximants. Using state-of-the-art models, we apply this new technique to the binary neutron star observations GW170817 and GW190425 and \changedtwo{to} the sub-threshold candidate GW200311\_103121. \changedtwo{Our analysis reveals subtle systematic differences (with Bayesian odds of $\sim 2$) between waveform models. A frequency-dependence study suggests that this may be due to the treatment of the tidal sector.} This new technique provides a proving ground for model development and a means to identify waveform-systematics in future observing runs where detector improvements will increase the number and clarity of binary neutron star collisions we observe.
\end{abstract}

\maketitle

\section{Challenges in gravitational-wave modelling}
\label{sec:challenges}

The first detection of gravitational waves and electromagnetic signals originating from the same astrophysical source, the merger of two neutron stars \seventeennoxspace\cite{GW170817_discovery}, revolutionised astronomy and led to advances in numerous scientific fields, e.g., a new and independent way to measure the Hubble constant\cite{Abbott:2017xzu,Hotokezaka:2018dfi,Dietrich:2020lps}, the proof that neutron star mergers are a cosmic source of heavy elements\cite{Cowperthwaite:2017dyu,Smartt:2017fuw,Kasliwal:2017ngb,Kasen:2017sxr}, tight constraints on alternative theories of gravity\cite{Ezquiaga:2017ekz,Baker:2017hug,Creminelli:2017sry}, and a measurement of the propagation speed of gravitational waves\cite{GBM:2017lvd}. 
Since this breakthrough detection, the LIGO-Scientific and Virgo Collaborations have observed a second binary neutron star merger \nineteennoxspace\cite{GW190425} and reported the sub-threshold candidate \subthresnoxspace\cite{GWTC3}.

Astrophysical inferences about gravitational-wave events rely on an accurate measurement of the source properties, e.g., the mass and spin of the component stars, the luminosity distance, and the tidal properties. 
Measuring these properties is part of the \emph{inverse problem}.
Typically, a Bayesian inference approach is applied, which requires \order{10^8} model evaluations to robustly infer the \emph{posterior distribution} for the tens of parameters that describe a binary neutron star merger.
Given the complexity of Einstein's Field Equations, which govern the final stages of the coalescence, the direct computation of gravitational waveforms is a challenging task.
State-of-the-art numerical-relativity simulations, in which the equations of general relativity and general-relativistic hydrodynamics are solved, require considerable resources on high-performance computing centres to model the dynamics and gravitational-signal emitted shortly before the merger of the two neutron stars. 
Despite their need for millions of CPU hours, these simulations allow only the study of the last 10 to 20 orbits before the collision\cite{Bruegmann:2018,Dietrich:2018phi,Kiuchi:2019kzt}.
On the other hand, the Advanced LIGO\cite{LIGOScientific:2014pky} and Virgo\cite{VIRGO:2014yos} detectors have a broadband sensitivity which enables them to measure several thousand orbits before the merger.
Given these restrictions, the direct computation of gravitational-waveform models to solve the inverse problem for binary merger events is impossible.

Therefore, the analysis of gravitational-wave signals of binary neutron star systems relies on the usage of analytical and semi-analytical \emph{approximant models}.
These approximant models are either based on the Post-Newtonian (PN) framework\cite{Blanchet:2013haa}, a perturbative approach to solve Einstein's Field equations for small velocities and large distances, on the effective-one-body (EOB) approach\cite{Buonanno:1998gg,Buonanno:2000ef}, in which the relativistic two-body problem is mapped into an effective one-body description, or simplified phenomenological models that incorporate PN knowledge and are calibrated through EOB and numerical-relativity data\cite{Dietrich:2017aum}. 
With these approaches, the approximant models can be evaluated in a few tens of milliseconds, enabling the source properties to be inferred in a few days.

The gravitational-wave community has made significant progress in improving these waveform approximants over the last few years. Higher tidal PN contributions have been computed\cite{Blanchet:2013haa}, different tidal EOB approximants have been developed\cite{Bernuzzi:2014owa,Hotokezaka:2015xka,Hinderer:2016eia}, and numerous phenomenological models have been derived\cite{Dietrich:2017aum,Dietrich:2019kaq,Kawaguchi:2018gvj}. 
To this extent, the reliability of waveform approximants was always checked against numerical-relativity simulations, which introduces additional challenges. First, the error assessment of general-relativistic hydrodynamics simulations is complicated due to the formation of shocks and discontinuities in the matter fields. Second, the simulations can only cover the late inspiral. Therefore, although there have been works that showed possible waveform systematic biases for future detections\cite{Dudi:2018jzn,Samajdar:2018dcx,Gamba:2020wgg,Pratten:2021pro,Kunert:2021hgm}, a qualitative judgment about the accuracy of the waveform models has always been difficult.

The standard approach to account for intrinsic modelling errors is to study differences between the inferred posterior distribution for a set of approximant models. Then, these differences in the posteriors are investigated through the direct computation of a few numerical-relativity waveforms in the problematic parameter space region with the goal to understand if the differences point to a deficiency of one or more of the models.
If the models hold up under investigation, the differences are ascribed to ``waveform systematics''.
To produce posterior distributions, which account for waveform systematics, it is usual to mix together the posterior distribution from different approximants.
This process yields a posterior distribution marginalised over the uncertainty inherent in the predefined set of models.
Typically\cite{GWTC3}, this is done by mixing together the equal-weighted posteriors from each model.
However, an equal-weighted approach neglects information provided by the Bayesian evidence; it is instead preferable to mix posteriors according to their \changedtwo{relative Bayesian evidence}\cite{Ashton:2019leq}.
This process also provides a means to study model selection using the Bayesian odds between models.
However, the previous approach\cite{Ashton:2019leq} suffers two difficulties.
\changedtwo{First, it relies on estimation of the Bayesian evidence and uncertainty using, for example, the Nested Sampling algorithm\cite{Skilling:2006gxv}.
The robustness of such a result can be difficult to guarantee in practice and, as we show in Section~\ref{sec:method}, the method developed herein can achieve more precise measurement of the odds at nearly equal computational expense compared to the evidence-weighting approach.
}
Second, from a pragmatic point of view, it is sometimes problematic to ensure independent analyses are identical in all respects except the model approximant. This is because slight differences in, for example, the analysed data systematically impact the Bayesian evidence and can result in systematic errors which are difficult to identify.

In this work \changedtwo{(see Section~\ref{sec:method})}, we develop a new data-driven validation of gravitational-wave approximants using the idea of \emph{hypermodels}: simultaneously inferring the source properties of the event by applying stochastic sampling to a predefined set of waveform models.
Building on a similar grid-based approach\cite{Jan:2020bdz}, this technique can produce posterior inferences directly marginalising over the hypermodel set, capturing the intrinsic modelling uncertainty\footnote{\changedtwo{The approach discussed herein parallels the grid-based approach\cite{Jan:2020bdz} in averaging over models in the likelihood, rather than weighting by the evidence. However, it improves the approach by including and marginalizing over the posterior model weights. This enables the algorithm to measure the fit of each waveform model and provide a set of posteriors weighted by their relative fit.}} 
But, it also enables inferences about the predictions of individual models in the hypermodel set and the plausibility of each model relative to all other models in the set.
This allows us to understand which of the existing waveform models is preferred and describes the observational data best.
By varying the choice of frequency-domain data, we can show how the method can reveal in which frequency range noticeable differences between the models occur, which leads to insights about the models themselves.

\section{Inferring information from gravitational-wave data}
\label{sec:inferring}

We apply our new method to study the first two confidently detected binary neutron star signals observed by the Advanced LIGO and Advanced Virgo detectors, \seventeennoxspace\cite{GW170817_discovery} and \nineteennoxspace\cite{GW190425}, and the
sub-threshold candidate \subthres recently reported in the GWTC3 catalogue\cite{GWTC3}.
We analyse the data with four cutting-edge spin-aligned waveform approximant models for binary neutron star mergers: \imrDNRTnoxspace\cite{Husa:2015iqa, Khan:2015jqa, Dietrich:2019kaq}, \seoNRTnoxspace\cite{Bohe:2016gbl, Dietrich:2019kaq}, \seoTnoxspace\cite{Hinderer:2016eia, Lackey:2018zvw}, and \teonoxspace\cite{Nagar:2018zoe}.
All four approximants, neglect the effects of precession (analyses of binary neutron star systems using precessing waveform models demonstrate the effect is negligible\cite{GW170817_properties, GW190425}), but include matter-effects through the \emph{tidal parameters}.

For each event, we analyse \SI{128}{s} of data covering the event and use the on-source Power Spectral Density computed by \bayesline~\cite{Littenberg:2014oda} and published with the original discovery.
For \seventeen and \subthres, we analyse the frequency-domain data from \SI{23}{Hz} to \SI{2048}{Hz}, while for \nineteen we analyse data from \SI{20}{Hz} to \SI{2048}{Hz} (the difference in the lower bound arises from the different total mass of the systems).
Unlike the original analyses, we exclude the marginalisation over the systematic error in the measured astrophysical strain due to the detector calibration.
This error is sub-dominant to the systematic errors from waveform modelling,\cite{Payne:2020myg} and we, therefore, neglect it.
We apply an astrophysically motivated\cite{TheLIGOScientific:2017qsa} low spin prior, restricting the dimensional spin magnitudes of each component to be less than 0.05, where this bound is derived from observed binary neutron star systems and theoretical spin estimates at their respective moment of the merger.
For other parameters, we use non-informative priors, i.e., uniform in the component masses with cuts made in the chirp mass and mass ratio and uninformative priors for all other parameters.\cite{Veitch:2014wba}
The exception to this is the analysis of \seventeen, in which we fix the sky location to that of the observed electromagnetic counterpart.

\begin{figure*}[t]
    \centering
    \includegraphics[width=0.95\textwidth]{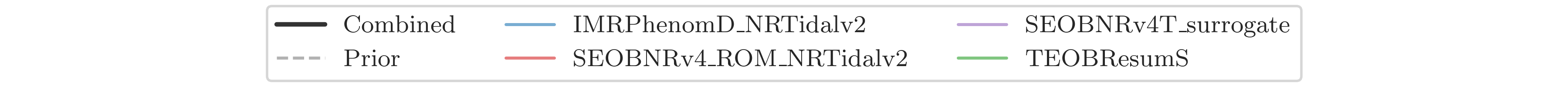}
    \includegraphics{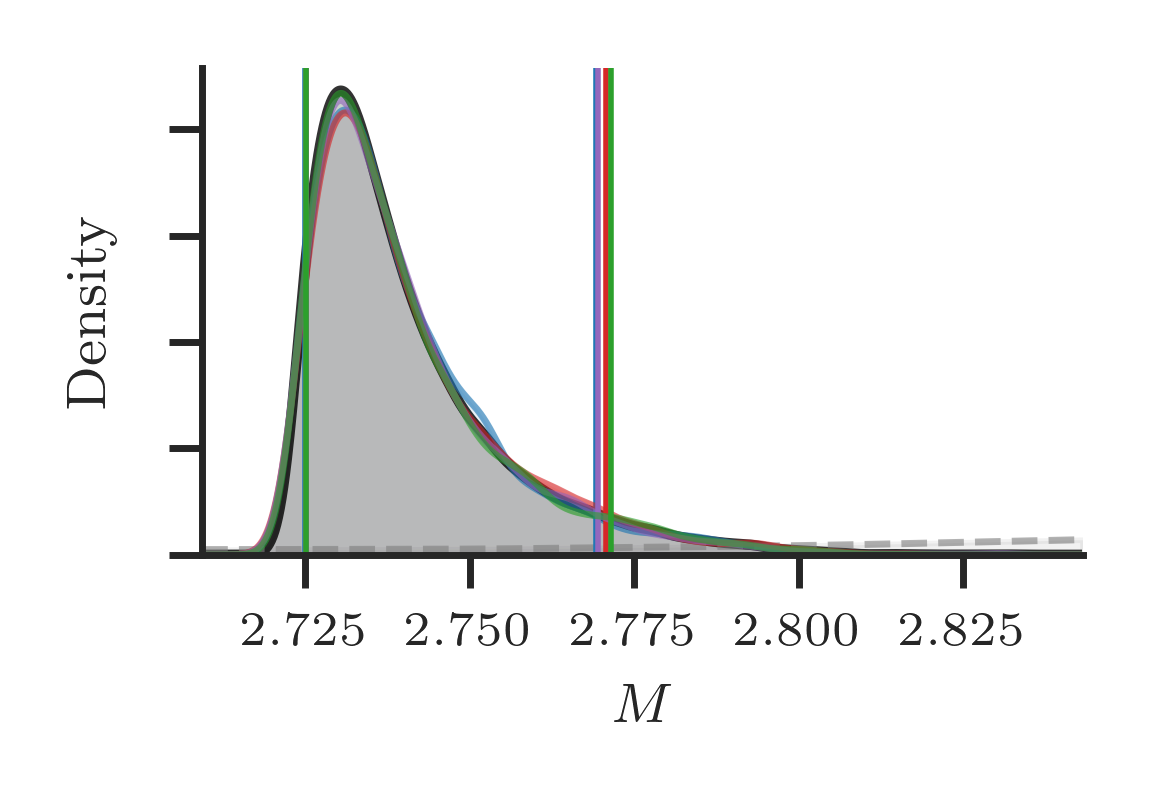}
    \includegraphics{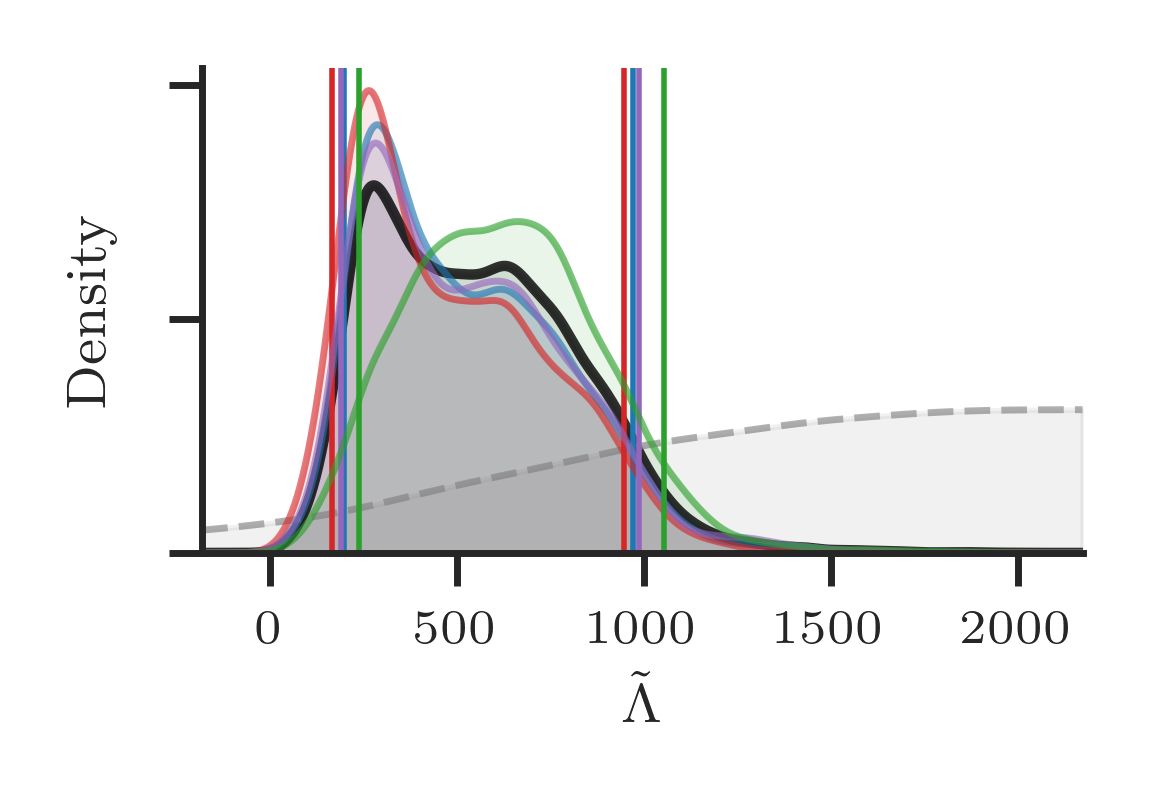}
    \includegraphics{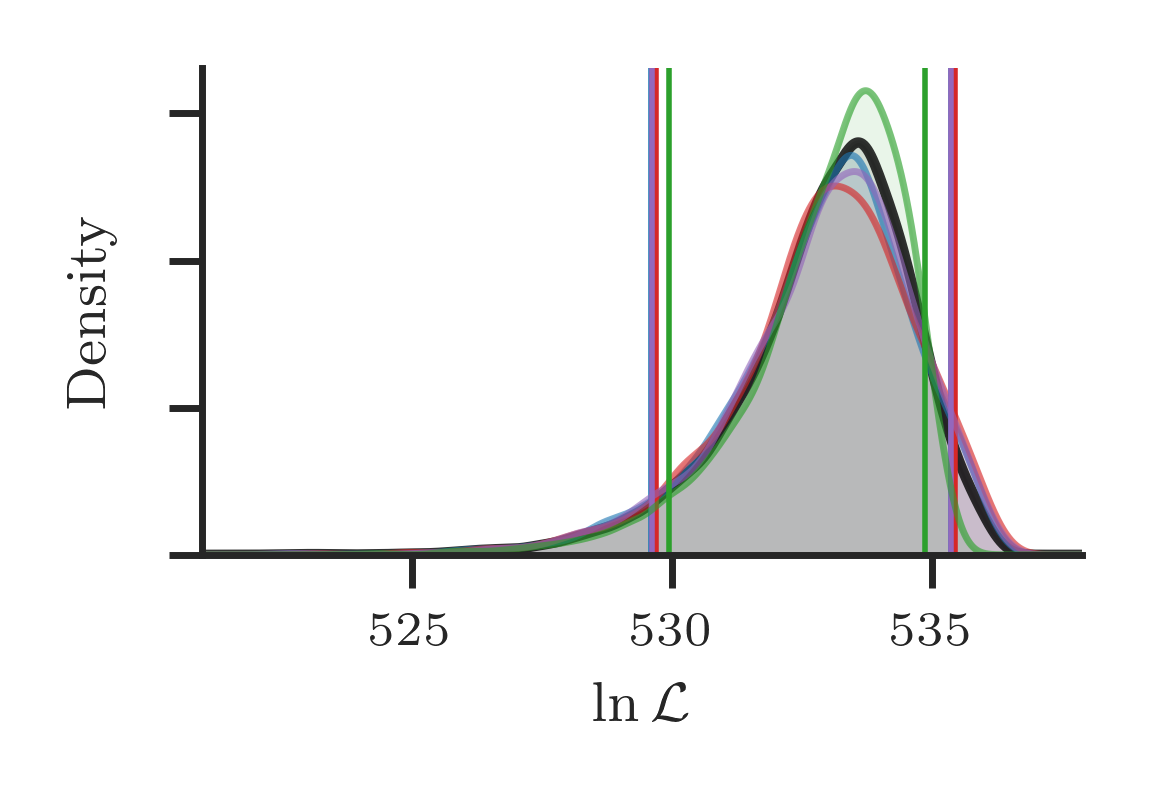}
    \includegraphics{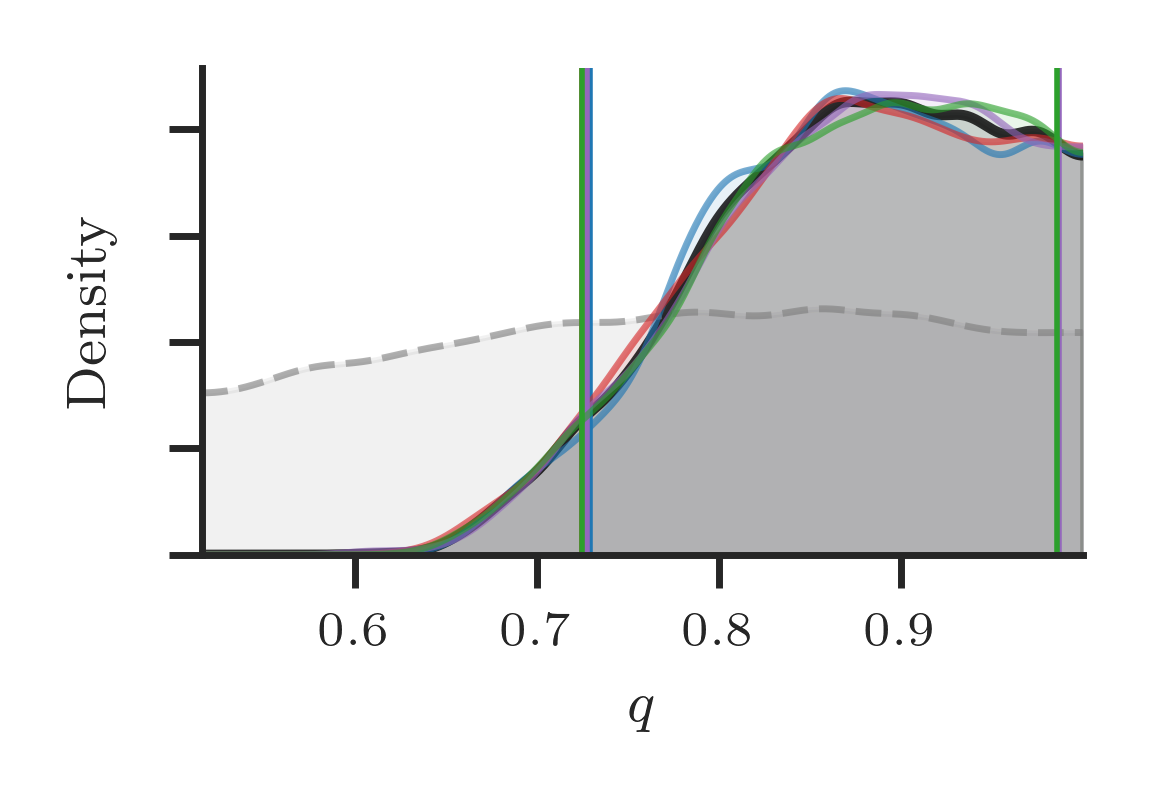}
    \includegraphics{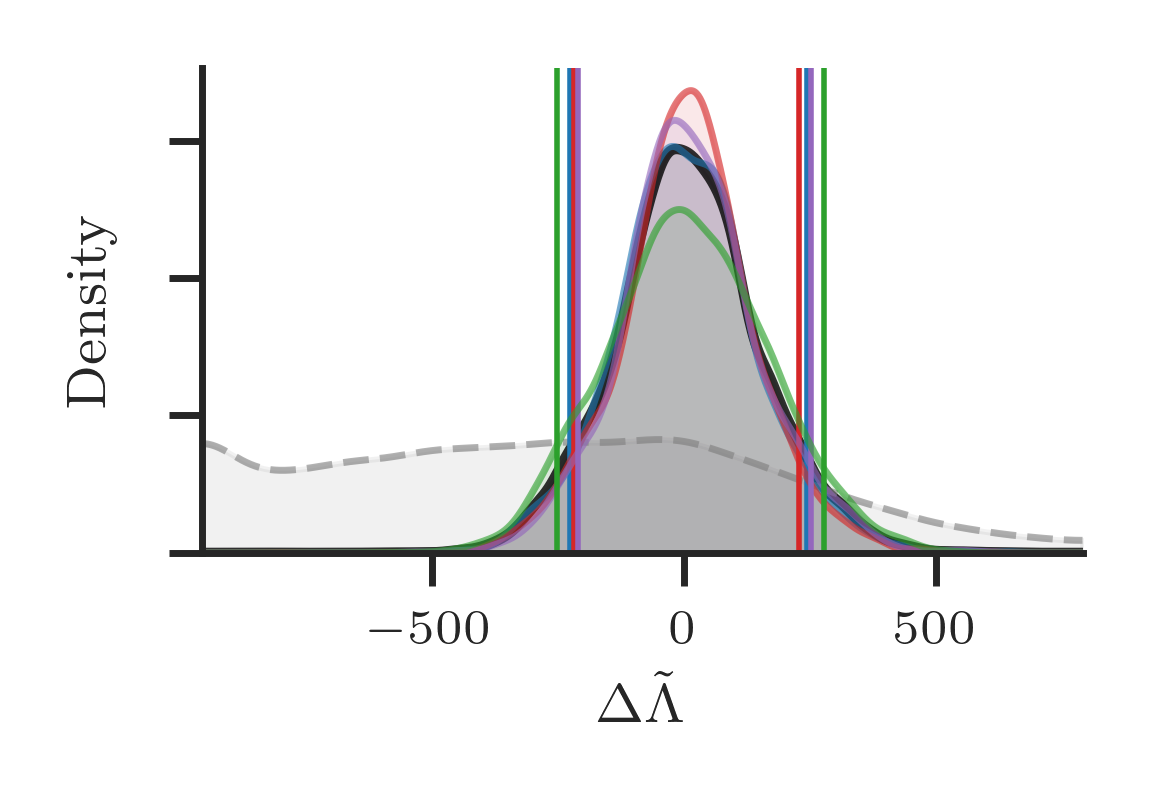}
    \includegraphics{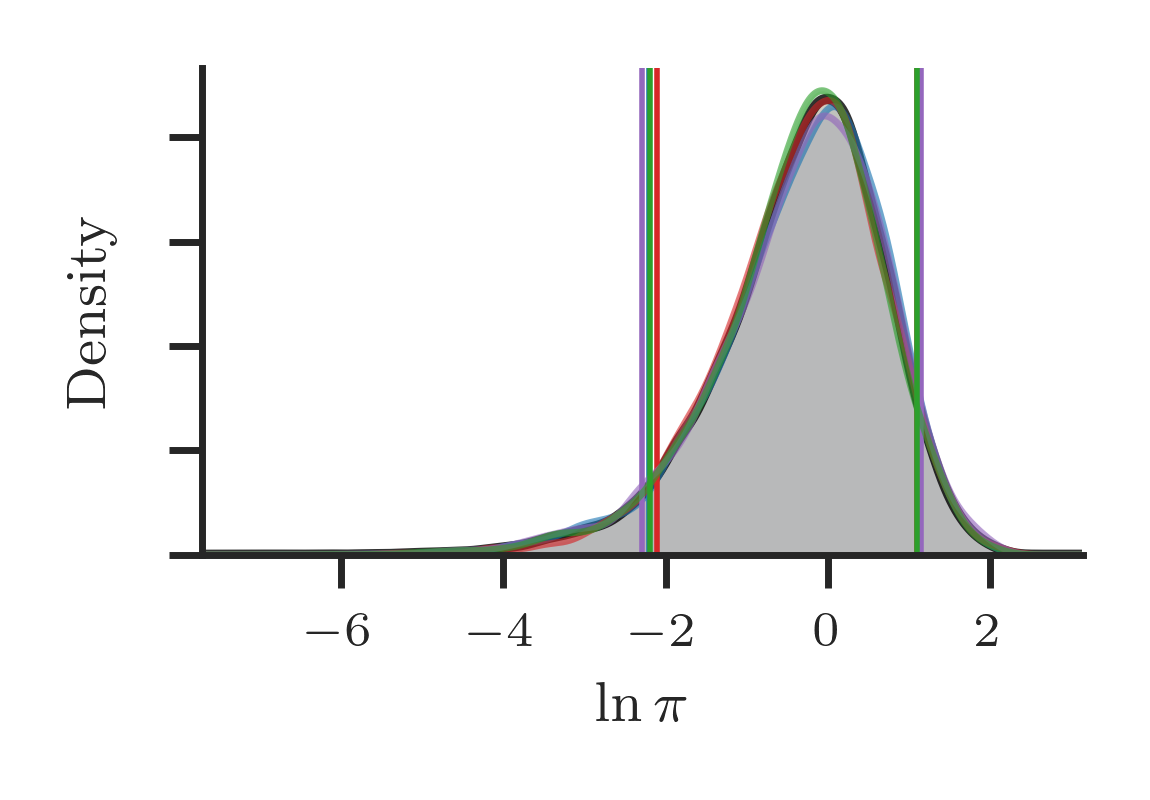}
    \caption{
    Posterior densities for the mass and  tidal deformability of \seventeen. We show the combined posteriors, marginalized over all waveform models using the hypermodel approach, and the posteriors for each individual model, extracted from the combined posterior, as well.
    The dashed curve provides the prior distribution estimated by drawing samples.
    In the right-hand column, we include the distributions of the log-likelihood and log-prior of the posterior samples.}
    \label{fig:GW170817}
\end{figure*}

\setlength{\tabcolsep}{3pt}
\begin{table*}[t]
    \centering
    \begin{tabular}{l|cc|cc|cc|c}
         \multirow{2}{*}{Waveform} &  \multicolumn{2}{c|}{\seventeen} & \multicolumn{2}{c|}{\nineteen} & \multicolumn{2}{c|}{\subthres} & Joint \\
         & Prob. [\%] & odds & Prob. [\%] & odds & Prob. [\%] & odds & odds \\\hline
         \imrDNRT
         & $\seventeenImrphenomdNrtidalvtwoPercentage \pm \seventeenImrphenomdNrtidalvtwoPercentageUncertainty$
         & $\seventeenMcmcOddsTeobresumsVsImrphenomdNrtidalvtwo \pm \seventeenMcmcOddsUncertaintyTeobresumsVsImrphenomdNrtidalvtwo$~  ($\seventeenDynestyOddsTeobresumsVsImrphenomd\pm\seventeenDynestyOddsUncertaintyTeobresumsVsImrphenomd$)
         & $\nineteenImrphenomdNrtidalvtwoPercentage \pm \nineteenImrphenomdNrtidalvtwoPercentageUncertainty$
         & $\nineteenMcmcOddsTeobresumsVsImrphenomdNrtidalvtwo\pm\nineteenMcmcOddsUncertaintyTeobresumsVsImrphenomdNrtidalvtwo$
         & $\subthresImrphenomdNrtidalvtwoPercentage \pm \subthresImrphenomdNrtidalvtwoPercentageUncertainty$
         & $\subthresMcmcOddsTeobresumsVsImrphenomdNrtidalvtwo\pm\subthresMcmcOddsUncertaintyTeobresumsVsImrphenomdNrtidalvtwo$
         & \fpeval{round(\seventeenMcmcOddsTeobresumsVsImrphenomdNrtidalvtwo*\nineteenMcmcOddsTeobresumsVsImrphenomdNrtidalvtwo*\subthresMcmcOddsTeobresumsVsImrphenomdNrtidalvtwo, 1)}
         $\pm$
         \fpeval{round(\seventeenMcmcOddsTeobresumsVsImrphenomdNrtidalvtwo*\nineteenMcmcOddsTeobresumsVsImrphenomdNrtidalvtwo*\subthresMcmcOddsTeobresumsVsImrphenomdNrtidalvtwo*sqrt(\seventeenMcmcOddsUncertaintyTeobresumsVsImrphenomdNrtidalvtwo**2 + \nineteenMcmcOddsUncertaintyTeobresumsVsImrphenomdNrtidalvtwo**2 +\subthresMcmcOddsUncertaintyTeobresumsVsImrphenomdNrtidalvtwo **2 ), 1)}
         \\
         \seoNRT 
         & $\seventeenSeobnrvfourRomNrtidalvtwoPercentage \pm \seventeenSeobnrvfourRomNrtidalvtwoPercentageUncertainty$
         & $\seventeenMcmcOddsTeobresumsVsSeobnrvfourRomNrtidalvtwo\pm\seventeenMcmcOddsUncertaintyTeobresumsVsSeobnrvfourRomNrtidalvtwo$~  ($\seventeenDynestyOddsTeobresumsVsSeobnrvfour\pm\seventeenDynestyOddsUncertaintyTeobresumsVsSeobnrvfour$)
         & $\nineteenSeobnrvfourRomNrtidalvtwoPercentage \pm \nineteenSeobnrvfourRomNrtidalvtwoPercentageUncertainty$
         & $\nineteenMcmcOddsTeobresumsVsSeobnrvfourRomNrtidalvtwo\pm\nineteenMcmcOddsUncertaintyTeobresumsVsSeobnrvfourRomNrtidalvtwo$
         & $\subthresSeobnrvfourRomNrtidalvtwoPercentage \pm \subthresSeobnrvfourRomNrtidalvtwoPercentageUncertainty$
         & $\subthresMcmcOddsTeobresumsVsSeobnrvfourRomNrtidalvtwo\pm\subthresMcmcOddsUncertaintyTeobresumsVsSeobnrvfourRomNrtidalvtwo$
         & \fpeval{round(\seventeenMcmcOddsTeobresumsVsSeobnrvfourRomNrtidalvtwo*\nineteenMcmcOddsTeobresumsVsSeobnrvfourRomNrtidalvtwo*\subthresMcmcOddsTeobresumsVsSeobnrvfourRomNrtidalvtwo, 1)}
         $\pm$
         \fpeval{round(\seventeenMcmcOddsTeobresumsVsSeobnrvfourRomNrtidalvtwo*\nineteenMcmcOddsTeobresumsVsSeobnrvfourRomNrtidalvtwo*\subthresMcmcOddsTeobresumsVsSeobnrvfourRomNrtidalvtwo*sqrt(\seventeenMcmcOddsUncertaintyTeobresumsVsSeobnrvfourRomNrtidalvtwo**2 + \nineteenMcmcOddsUncertaintyTeobresumsVsSeobnrvfourRomNrtidalvtwo**2 +\subthresMcmcOddsUncertaintyTeobresumsVsSeobnrvfourRomNrtidalvtwo **2 ), 1)}
         \\
         \seoT
         & $\seventeenSeobnrvfourtSurrogatePercentage \pm \seventeenSeobnrvfourtSurrogatePercentageUncertainty$ 
         & $\seventeenMcmcOddsTeobresumsVsSeobnrvfourtSurrogate\pm\seventeenMcmcOddsUncertaintyTeobresumsVsSeobnrvfourtSurrogate$~  ($\seventeenDynestyOddsTeobresumsVsSeobnrvfourt\pm\seventeenDynestyOddsUncertaintyTeobresumsVsSeobnrvfourt$)
         & $\nineteenSeobnrvfourtSurrogatePercentage \pm \nineteenSeobnrvfourtSurrogatePercentageUncertainty$
         & $\nineteenMcmcOddsTeobresumsVsSeobnrvfourtSurrogate\pm\nineteenMcmcOddsUncertaintyTeobresumsVsSeobnrvfourtSurrogate$
         & $\subthresSeobnrvfourtSurrogatePercentage \pm \subthresSeobnrvfourtSurrogatePercentageUncertainty$
         & $\subthresMcmcOddsTeobresumsVsSeobnrvfourtSurrogate\pm\subthresMcmcOddsUncertaintyTeobresumsVsSeobnrvfourtSurrogate$
         & \fpeval{round(\seventeenMcmcOddsTeobresumsVsSeobnrvfourtSurrogate*\nineteenMcmcOddsTeobresumsVsSeobnrvfourtSurrogate*\subthresMcmcOddsTeobresumsVsSeobnrvfourtSurrogate, 1)}
         $\pm$
         \fpeval{round(\seventeenMcmcOddsTeobresumsVsSeobnrvfourtSurrogate*\nineteenMcmcOddsTeobresumsVsSeobnrvfourtSurrogate*\subthresMcmcOddsTeobresumsVsSeobnrvfourtSurrogate*sqrt(\seventeenMcmcOddsUncertaintyTeobresumsVsSeobnrvfourtSurrogate**2 + \nineteenMcmcOddsUncertaintyTeobresumsVsSeobnrvfourtSurrogate**2 +\subthresMcmcOddsUncertaintyTeobresumsVsSeobnrvfourtSurrogate **2 ), 1)}
         \\
         \teo
         & $\seventeenTeobresumsPercentage \pm \seventeenTeobresumsPercentageUncertainty$ 
         & ---
         & $\nineteenTeobresumsPercentage \pm \nineteenTeobresumsPercentageUncertainty$
         & ---
         & $\subthresTeobresumsPercentage \pm \subthresTeobresumsPercentageUncertainty$
         & ---
         & ---
         \\
    \end{tabular}
    \caption{The posterior probability, as a percentage, for each waveform and data set analysed in this work.
    Next to the probability, we also provide the odds against the \teo waveform (calculated from the ratio of the posterior probability).
    For \seventeen, we also give the odds calculated from a Nested Sampling approach in brackets.
    All uncertainties are stated as $1\sigma$ bounds.
    Uncertainties on the posterior probabilities are derived from Poisson statistics, while the uncertainties on the Nested Sampling odds are derived from estimates reported by the \dynesty algorithm.
    }
    \label{tab:posterior_prob}
\end{table*}

\subsection{\label{sec:GW170817}\seventeen:}
\seventeen was the first observation of a gravitational-wave signal emitted from a binary neutron star merger. 
Because of its small distance, \SI{40}{Mpc}, combined with its long duration, it is the observation with the largest signal-to-noise (SNR) detected so far.
This large SNR, 32.4, allows to extract source properties such as the total masses $M=M_A+M_B$, the mass ratio $q=M_A/M_B\leq 1$, and information about the star's deformability. 
Considering the latter, the finite size of the two stars and the deformations of the stars within the gravitational field of their companion creates a characteristic imprint into the waveform, which is distinct from that of a binary black hole.
Inferences of the stars deformability provide a unique probe of the properties of supranuclear dense matter.
These imprints are mainly characterized by the binaries tidal deformability\cite{Flanagan:2007ix}:
\begin{equation}
\tilde{\Lambda} = \frac{16}{13} 
\frac{(M_A+12M_B) M_A^4 \Lambda_2^A + (M_B+12M_A) M_B^4 \Lambda_2^B}{(M_A+M_B)^5},
\end{equation}
where $\Lambda^{A,B} =2/3k_2^{A,B} (c^2/G) R_{A,B}/M_{A,B}^5$ 
are the individual tidal deformabilities or polarizability 
with the second Love Number $k_2$, the stellar radius $R_{A,B}$, 
and the mass of the individual stars $M_{A,B}$. 

In \cref{tab:posterior_prob}, we give the posterior probability for each waveform, calculated from the fraction of posterior samples drawn from each waveform.
These posterior probabilities measure the relative success of the different waveform models at predicting the data (normalised by the finite set of models considered).
Of the four waveforms, the \teo waveform is the most successful at predicting the \seventeen data (\seventeenTeobresumsPercentage\% as compared to the next largest value \seventeenSeobnrvfourtSurrogatePercentage\% for \seoT).
Taking the ratio of posterior probabilities, we can convert the posterior probability into a Bayesian odds of \teo relative to the other models, which ranges from 1.4 to 1.6 (the odds are equivalent to the Bayes factor as we set equal prior odds between models).
These odds do not rise to the level of substantial evidence favouring the \teo model (see, e.g., the interpretation given by Ref.\cite{Kass:1995loi} which suggests a threshold of 3.2).
However, the mild preference for \teo is worthy of further investigation given the potentially drastic implications for future observations and the necessary development of modelling approaches.
\changedtwo{We discuss the interpretation of the odds further in Sec.~\ref{sec:supplementary}.}

To delve into why \teo may be preferred, in Fig.~\ref{fig:GW170817}, we plot the inferred posteriors of the total mass, mass ratio, tidal deformability, and \deltalambdatilde,  another mass-weighted combination of $\Lambda_{A, B}$ which characterises higher-order contributions\cite{Harry:2018hke}.
In each figure, we give the ``Combined'' result marginalised over the four waveform models and the separated posterior from each waveform model.
We find strong agreement between the four models for the intrinsic mass of the system, but moderate differences for \lambdatilde and \deltalambdatilde.\footnote{We also find strong agreement for all other intrinsic and extrinsic parameters of the system, details of which can be found in the data release}
The posterior distribution of \lambdatilde predicted by \teo supports larger values than the other three waveform models, while the \deltalambdatilde distribution is wider.
These findings replicate that of previous work\cite{Gamba:2020wgg}.
Our new result further demonstrates that this difference is accompanied by a mild preference for \teo over the other waveform models.

It is wise to consider why our algorithm finds a preference for \teo.
From the sampling perspective, information about the evidence for and against each waveform is contained solely in the distribution of log-prior and log-likelihood values, which we also visualise in \cref{fig:GW170817}.
Let us consider the log-prior distribution first: comparing the four waveform models, we do not observe any trends in the log-prior distribution.
This indicates that all of the information is arising from the log-likelihood distribution.
Turning to the log-likelihood distribution, we find that the \teo distribution contains a prominent peak compared to the other waveform models.
However, its maximum likelihood point and the 95\% quantile are smaller than the other waveform models.
So, it is preferred not because it has a larger maximum likelihood but rather because of the shape and location of the distribution of the likelihood, i.e., the median is at a larger log-likelihood value and the distribution is more tightly constrained to higher values.
This underlines the inherent danger of a maximum likelihood analysis which would conclude that \teo is the worst performing model.

To validate our results, we repeat our analyses using a Bayesian evidence approach. We analyse each event individually using the \dynesty Nested Sampling package to calculate the Bayesian evidence.
In \cref{tab:posterior_prob}, we report the odds of each model against \teo for \seventeen in brackets to show that the odds, as calculated from a Nested Sampling approach, agree with our hypermodel approach to within the stated uncertainties.
The uncertainty on the odds calculated from the Nested Sampling approach is larger than that of the hypermodels approach.
The reason for this is explained in \cref{sec:method}, but we note here that, while we can reduce the uncertainty in either approach by additional computation effort, the uncertainty of the hypermodel approach is minimised for nearly equally favoured models, making it well suited to problems such as this.

Finally, we note that the \teo model can include the impacts of higher-order mode waveform content.
For the primary analyses in this work, we restricted the \teo to only model the $\ell=2, m=\pm2$ mode (all other waveform approximants only model this mode).
To explore if higher-order mode content is measurable in \seventeen, we repeat our Nested Sampling analysis (using a massively-parallelised approach) for the \teo waveform, but include all modes up to the $\ell=4, m=\pm4$ harmonics.
We then compare the posterior and Bayesian evidence between the analysis with and without higher-order modes and find they are identical, i.e., we do not find any evidence for higher-order modes in \seventeen.
This is expected: for systems with near-equal component masses, less than $0.2\%$ of the total emitted gravitational-wave energy is released in higher-order modes\cite{Dietrich:2016hky}. 


\subsection{\label{sec:GW190425}\nineteen:}

Next, we analyse the second observed binary neutron star merger \nineteennoxspace\cite{GW190425}.
Unlike \seventeen, no electromagnetic counterpart was identified alongside \nineteen.
Moreover, the event had an SNR of only 13.
Therefore the data individually places weaker constraints on the tidal deformability (though it still does contribute some information\cite{GW190425}).

We apply our hypermodel analysis to \nineteen in a manner identical to our analysis of \seventeen (except that, without an electromagnetic counterpart, we must include the prior uncertainty about the sky position).
In \cref{tab:posterior_prob}, we provide the posterior probability.
Remarkably, we find a consistent pattern emerging: \teo is the most successful model at predicting the data.
The ranking of the other three waveforms is nearly the same, \seoNRT ranks last with \imrDNRT and \seoT comparable to within their stated uncertainties (though their ordering is flipped as compared to \seventeen).

Like \seventeen, and in agreement with previous analyses\cite{GW190425}, all four waveform models predict identical posteriors for all parameters except \lambdatilde (see \cref{sec:posterior} for additional figures).
For \lambdatilde we find a subtle difference in predictions for \teo compared to the other waveforms.

Comparing \nineteen and \seventeen, we obtain consistent but weaker inferences about the probability of the four waveforms and inferences of the tidal parameters, which is expected since \nineteen is an intrinsically quieter source.

\subsection{\label{sec:S200311ba}\subthres:}
Finally, we analyse the sub-threshold binary neutron star candidate event \subthresnoxspace\cite{GWTC3}.
This candidate has an SNR of only $\sim 9$.
Analysing the event under the assumption that it is an astrophysical signal, its properties are highly consistent with that of a binary neutron star merger.
However, the probability that this candidate is astrophysical is estimated to be 19\% by the PyCBC-broad search, but just 3\% by the MBTA search.
Hence, even with the most optimistic estimate of its astrophysical probability, current analyses conclude the candidate is more likely to be
non-astrophysical than a real signal.
Nevertheless, for completeness, we choose to analyse the event using our hypermodel approach (results are presented in \cref{sec:posterior} of the Supplementary Material). Moreover, \subthres provides a helpful test to understand how indeterminate candidates can affect our analysis.

Our combined analysis predicts component masses of $1.4^{+0.2}_{-0.2}$ and $1.2^{+0.1}_{-0.2}$, placing them in the centre of the theoretically predicted distribution of such systems.
However, due to the low SNR of the system, all other physical parameters are essentially unconstrained.
The tidal parameters follow the prior distributions, and we do not observe a preference for \teo or any other waveform from the study of waveform models.
This is unsurprising given the low SNR of the event.
In \cref{tab:posterior_prob}, we combine the odds from all three events.
Formally, we are neglecting the information that \subthres may not be an astrophysical signal. 
However, because the inferences the data provide on the relative likelihood of the four models is uninformative, combining it in this way does not produce a bias in the joint odds.

\section{Implication for gravitational-wave modelling}
\label{sec:implications}

Based on our findings, particularly the combined posterior probability in Tab.~\ref{tab:posterior_prob}, 
we find clear evidence that \teo explains both \seventeen and \nineteen marginally better than other models, but more data will be needed to verify this observation.
The sub-threshold candidate \subthres provides no additional constraints for or against \teo within the stated uncertainties.
The difference between \teo and the other models is modest: combining the odds from all three events together, the joint odds in the final column of \cref{tab:posterior_prob} lead to noticeable but not substantial evidence (see Sec.~\ref{sec:supplementary}).
Overall, our results are intriguing, especially given the consistency between two independent observations.

To interpret our result, we now review the differences between the four models: 
(i) \imrDNRT and \seoNRT use identical tidal contributions 
but different underlying point-particle baselines. \imrDNRT is marginally preferred over \seoNRT which suggests the underlying point-particle description predicts the data slightly better than \seoNRT;
(ii) \seoT and \seoNRT use almost identical point-particle descriptions but different tidal contributions,
hence, Tab.~\ref{tab:posterior_prob} reveals that the tidal description within \seoT is more accurate than the current NRTidalv2 version.
This is not surprising as \seoT contains, for example, non-adiabatic tidal effects. 
(iii) \teo, uses a point-particle and tidal contribution 
that is different from all other waveform models. 

\begin{figure}[t]
    \centering
    \includegraphics[width=0.95\columnwidth]{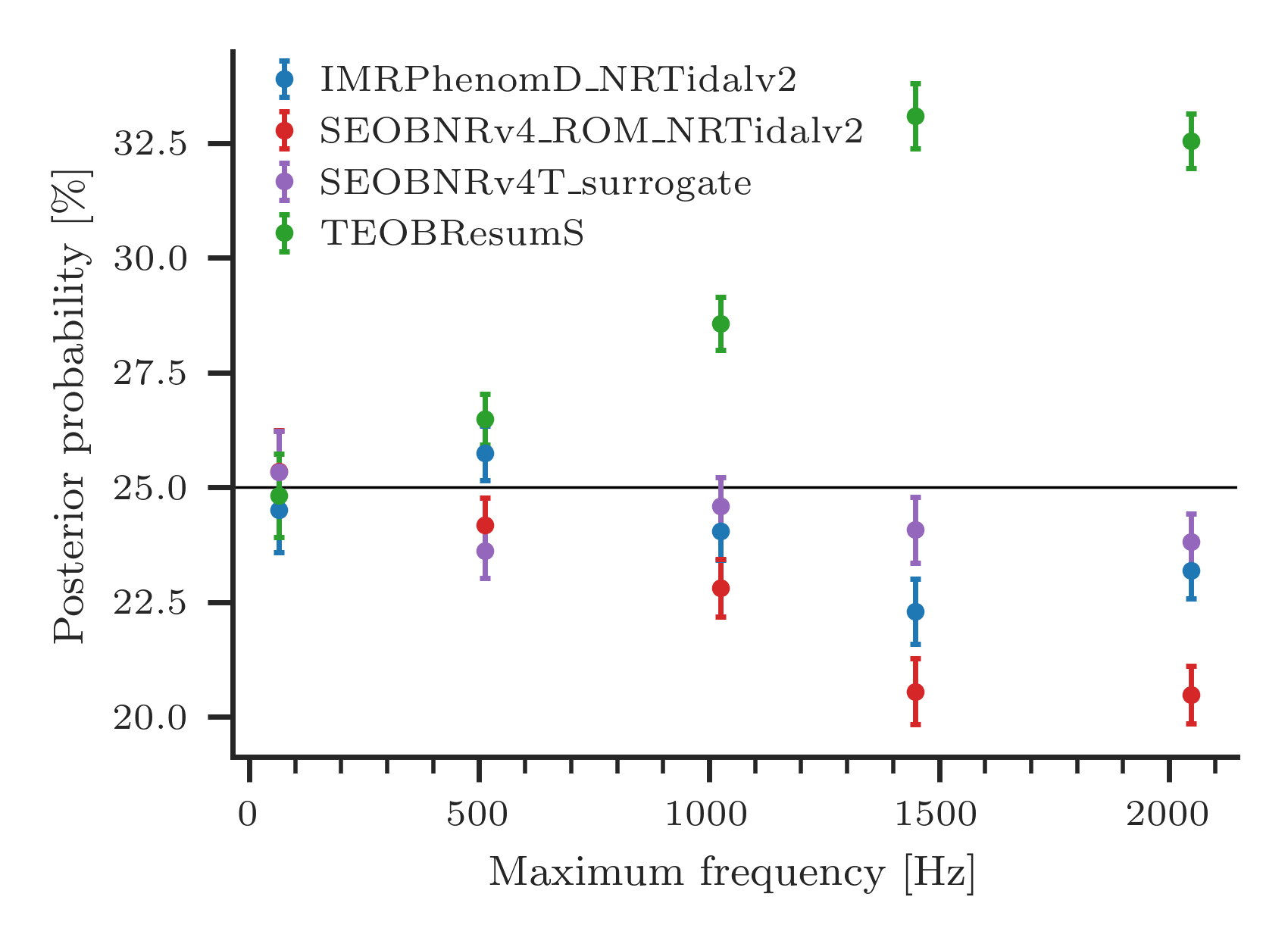}
    \caption{The evolution of the posterior probability of each waveform for \seventeen as the maximum frequency of the analysis data is varied.}
    \label{fig:fmax}
\end{figure}

To further investigate the result, in \cref{fig:fmax}, we repeat our analysis of \seventeen, but vary the maximum frequency of the analysis data.
This demonstrates that the evidence in support of \teo arises predominantly from 
the high frequency data (above \SI{512}{Hz}).
It is not possible to confidently identify which aspect of \teo is dominantly responsible for the difference.
However, given the high-frequency dependence of the effect, we note two potential reasons.
First, \imrDNRT, \seoNRT, and \seoT all employ a high-frequency waveform tapering while \teo does not.
To understand if this tapering causes the difference, we re-calculate the log-likelihood of the \imrDNRT samples for \seventeen using a modified model which excludes the tapering effect. The resulting distribution of log-likelihoods is statistically identical to the non-modified distribution. Hence, we conclude that tapering does not explain the difference.
Second, that the difference occurs predominantly in the frequency regime where tidal effects come to dominate, thus, we reason that it is the tidal sector that explains the effect, most likely the gravitational-self force inspired resummation of tidal potential present in \teo. 

\section{Conclusion}
\label{sec:conclusions}

In this work, we present a hypermodel approach to analysing binary neutron star mergers which 
    (i) provides ``on the fly'' marginalised posteriors distribution for gravitational-wave studies that reduce potential systematic 
    effects from gravitational-wave models;
    (ii) allows for model selection of gravitational-wave approximants; 
    (iii) tests gravitational-wave model assumptions without computationally expensive numerical-relativity simulations. 
We apply this approach to the two confidently detected binary neutron star collisions, \seventeen and \nineteen and the sub-threshold candidate \subthres.
We find a consistent preference for the \teo waveform model with an overall odds that ranges from \fpeval{round(\seventeenMcmcOddsTeobresumsVsSeobnrvfourRomNrtidalvtwo*\nineteenMcmcOddsTeobresumsVsSeobnrvfourRomNrtidalvtwo*\subthresMcmcOddsTeobresumsVsSeobnrvfourRomNrtidalvtwo, 1)} against \seoNRT to \fpeval{round(\seventeenMcmcOddsTeobresumsVsImrphenomdNrtidalvtwo*\nineteenMcmcOddsTeobresumsVsImrphenomdNrtidalvtwo*\subthresMcmcOddsTeobresumsVsImrphenomdNrtidalvtwo, 1)} against \imrDNRT
These odds fall short of substantial evidence, but the consistency between the events suggests that \teo is subtly better at explaining the observed data.
Identifying such a subtlety is essential.
Future observing runs of the LIGO, Virgo, and KAGRA detectors will be more sensitive thanks to developments in instrumentation.
This sensitivity translates into a greater clarity with which we observe events and hence, improved constraints on fundamental physics along with an increase in the number of detections\cite{KAGRA:2020npa}.
Therefore, future data will be critical to determine if \teo is better at predicting the data. 
However, we anticipate that all models considered herein will be further improved before the next observing run.
We encourage the waveform modelling community to validate new developments by rerunning the analyses in this work, potentially also by using numerical-relativity based injection data.

\section{Method}
\label{sec:method}

The source properties of gravitational-wave signals observed by ground-based interferometers are inferred using stochastic sampling\cite{Veitch:2014wba, GWTC1, GWTC2, GWTC2-1, GWTC3}.
Applying a Bayesian approach, the goal of sampling is to to approximate the posterior probability distribution
\begin{align}
    p(\params | \data, \model) \propto \likelihood(\data | \params, \model) \prior(\params | \model)\,,
\end{align} where $\params$ is a vector of the model parameters (e.g.,\ the mass and spin of the binary components), $\data$ is the time series of strain data recorded by the interferometer, $\model$ is the waveform approximant, \likelihood is the likelihood of the data given $\model$ and $\theta$, and \prior is the prior probability density for $\theta$ given $\model$.
Typically, a stochastic sampler produces an approximation of the posterior by generating a set of independent samples $\{\theta_i\}$ drawn from the posterior, which can be used, e.g., to calculate summary statistics.

In addition to the posterior, stochastic sampling can also approximate the Bayesian evidence $\evidence(\data | \model)$ which is fundamental to the notion of \emph{model comparison}.
Given multiple waveform approximant models, say $\model_A$ and $\model_B$,
robust measurements of the evidence can enable a model comparison via the \emph{Bayesian odds}:
\begin{equation}
    \underbrace{\odds_{A/B}}_{P(\model_A| \data)/P(\model_B | \data)} = \underbrace{\bayesfactor_{A/B}}_{\evidence(\data | \model_A)/\evidence(\data | \model_B)}
    \times
    \underbrace{\pi_{A/B}}_{\pi(\model_A)/\pi(\model_B)}\,.
    \label{eqn:odds}
\end{equation}
The odds $\odds_{A/B}$ are the relative probability of two models given the data and are calculated from the product of the data-driven \emph{Bayes factor} $\bayesfactor_{A/B}$, and the prior-odds $\pi_{A/B}$.
Typically, we have no prior preference between models such that $\pi_{A/B}=1$ and the odds and Bayes factor are identical.

Two stochastic sampling approaches have been demonstrated\cite{Christensen:1998gf, Veitch:2008ur} to be capable of robustly inferring both the posterior distribution and evidence of a gravitational-wave signal: Markov-Chain Monte-Carlo (MCMC)\cite{Metropolis:1953am, Hastings:1970aa} and Nested Sampling\cite{Skilling:2006gxv}.
In addition to MCMC and Nested Sampling, there are also grid-based approaches~\cite{Pankow:2015cra, Lange:2018pyp} which employ massive parallelisation and iterative fitting.
With appropriate tuning, both MCMC and Nested Sampling algorithms are roughly equally capable of approximating the posterior density.
However, Nested Sampling is more efficient in calculating the Bayesian evidence\cite{Veitch:2014wba, Ashton:2021anp}.
Therefore, the Nested Sampling approach is typically favoured in model selection problems.

However, calculating the odds between two models is only part of the problem.
Typically, several models are available, and often they are nearly equally favoured when confronted with observations\cite{Gamba:2020wgg, Estelles:2021jnz, Mateu-Lucena:2021siq, Colleoni:2020tgc}.
As a result, when drawing astrophysical conclusions from an event, it is essential to capture the systematic modelling uncertainty to avoid biased inferences.
Results published so far by the LIGO-Scientific and Virgo Collaborations
have addressed this issue by combining equal numbers of samples from a subset of pre-selected waveform models\cite{GWTC1, GWTC2, GWTC2-1, GWTC3}.
In effect, this presupposes that all waveform models are equally successfully at predicting the data.
However, this is certainly not the case and hence discards information about how well each model predicts the data.
An alternative approach\cite{Ashton:2019leq}, demonstrated how the Bayesian evidence can be used to capture this additional information, weighting samples and producing a set of posterior samples that marginalises over the pre-selected waveform models.
Subsequently, it was demonstrated\cite{Jan:2020bdz} how the grid-based approaches\cite{Lange:2018pyp} could be extended to perform model comparisons between pre-selected waveforms.
However, rather than calculating the Bayesian evidence, this approach instead included multiple models in the likelihood itself.

In this work, we introduce a new approach to calculating the odds between waveform approximants and calculating posteriors marginalised over a set of $n$ waveform models.
First, we extend the definition of the traditional waveform model to a hypermodel $\model\rightarrow \Omega=\{\model_{0}, \model_{1},\ldots \model_{n-1}\}$; when sampling we then infer the properties of the parameter set $\{\params, \omega\}$ where $\params$ is the usual vector of astrophysical model parameters while $\omega$ is
a categorical waveform-approximant parameter $\omega\in[0, 1, 2, \ldots, n-1]$.
We apply an uninformative prior on $\omega$, $\pi(\omega) = 1/n$.
Then, at each iteration of the MCMC sampler, following the standard Metropolis-Hastings algorithm\cite{Metropolis:1953am,Hastings:1970aa}, a new point in the set of model parameters is proposed, including a proposal for the categorical waveform model.
Given the proposed point, we first apply a pre-determined mapping between $\omega$ and the set of waveform approximants under study to select which waveform to use. Then, the likelihood is calculated according to the remaining model parameters.
We implement the categorical approach in the \bilbyMCMC sampler.
The only additional step is the addition of a specialised proposal routine for the categorical variable $\omega$, here we use a random draw from the prior.

It is worth pointing out that the MCMC-hypermodel approach does not add any additional computational cost compared to separate MCMC analyses.
One can quantify this by looking at the autocorrelation time $\tau$, which determines the number of steps required for the MCMC to produce a fixed number of independent samples\cite{Hogg:2017akh}.
Comparing an MCMC-hypermodel analysis with four waveforms to an identical analysis with a single waveform, we find that $\tau$ only differs by the expected uncertainty due to sampling, i.e., the addition of the hypermodel parameter does not increase $\tau$.
This is because $\tau$ is taken as the maximum over all parameters;
for the analyses presented here, the autocorrelation time of difficult-to-sample-parameters, such as the mass ratio, dominate, while the hypermodel parameter $\omega$ demonstrates good mixing and hence does not increase the maximum $\tau$.
The computational cost of the analysis is directly proportional to $\tau$\cite{Ashton:2021anp}.
Therefore, the total computational cost of running the four waveforms separately is nearly four times larger then running them together in the hyper-model approach if one only requires the same number of combined (waveform-marginalized) samples for the hypermodel approach compared to each individual run.
However, this is not a fair comparison, since one only obtains a quarter of the number of samples per waveform compared to running four individual analyses.
A fairer comparison is to require the same number of samples per waveform model.
Then, the two approaches have a near identical computational cost, with the hypermodel approach marginally more efficient because the MCMC burn-in cost is only paid once.
However, this does not hold if one model is strongly favoured.
In this case, the autocorrelation time of the hypermodel approach may be dominated by the waveform mixing and there are inefficiencies in running the sampler for a long duration to produce a fixed number of samples for the worst performing model.
For these reasons, and arguments below, we suggest a Nested Sampling approach instead be applied when one model is strongly preferred.

The MCMC-hypermodel approach applied in this work is a special case of the Reversible-Jump MCMC (RJMCMC) algorithm\cite{Green:1995mxx} which enables the models to differ in the model parameters.
In this work, we consider only models with identically-defined parameters \params. 
Hence, we distinguish our hypermodel approach from the RJMCMC algorithm.
However, in future work we hope to extend the sampling algorithm to a full RJMCMC sampler. This will enable the comparison of waveform models with differing nature, e.g., one could compare binary neutron star, neutron-star black-hole, and binary black hole models directly, or in comparing precessing and aligned-spin models.
The RJMCMC algorithm has been applied in other contexts in gravitational-wave data analysis before, e.g., for unmodelled searches\cite{Cornish:2007if, Cornish:2014kda}, power-spectral density estimation\cite{Littenberg:2014oda}, and population inferences\cite{Farr:2010tu}.

Running the MCMC-hypermodel sampler on $n$ waveform approximants, we obtain a set of $N$ independent samples from the posterior $\{\theta_{i}\}$.
By design, these samples are mixed according to the relative posterior probability of the individual waveform models.
Individual posterior distributions for each waveform approximant can be obtained by filtering the posterior against the relevant $\omega$ index.
The probability of the $\ell^{\rm}$-th waveform approximant relative to all other waveform approximants considered in the categorical analysis is $p_\ell=n_\ell / N$.
Finally, the Bayesian odds between two models ($\omega=A$ and $\omega=B$) is given by
\begin{align}
    \odds_{A/B} &= \frac{p_A}{p_B}\,,
\end{align}
with an estimated variance:
\begin{align}
    \sigma_{\odds_{A/B}}^{2} &\approx \frac{\odds_{A/B}^2}{N} \left(\frac{1}{p_A} + \frac{1}{p_B} \right)\,.
    \label{eqn:sigma}
\end{align}
If only two models are under consideration, this demonstrates that at leading order $\sigma_{\odds_{A/B}}^2\approx \odds_{A/B}^{2} / N$.
At fixed $N$, the variance has a minimum when $\odds_{A/B}=1$, i.e., the method is best-suited to study cases where $\odds \sim 1$, but the uncertainty grows with the odds.
For cases where the odds strongly favour one model over the other, $N$ can be increased to achieve a fixed level of uncertainty.
But, in cases where the odds are strongly informative (i.e., $\odds \gg 1$ or $\odds \ll 1$), a Nested Sampling approach will be more efficient.
For Nested Sampling, the evidence is estimated for each model individually and the uncertainty is independent of the odds.
Nested Sampling is therefore well suited to estimating odds for clear-cut model comparisons.
However, for cases where the models are in close contention, the scaling of the uncertainty with $N$ makes the MCMC approach preferable.
For example, while we use standard settings\cite{Ashton:2021anp} for both the MCMC and Nested Sampling methods, the uncertainties reported in Table~\ref{tab:posterior_prob} are nearly an order of magnitude larger for the Nested Sampling odds than that of the MCMC approach.
We could reduce the uncertainty on the Nested Sampling odds by changing the stopping criteria of the sampler with a corresponding increase in the computational cost.
While a detailed study of computational efficiency of the two approaches is beyond the scope of this analysis, we note here that the total number of CPU days was approximately 400 for the MCMC-hypermodel analysis of GW170817 while it was 350 for the Nested Sampling analysis illustrating that at a similar level of computing cost, the MCMC method outperforms the Nested Sampling approach when $\odds \sim 1$. (Note: all timing performed on Intel 2.4 GHz Gold 6148 CPUs).

The prior-odds, $\pi_{A/B}$, enter via the prior on the categorical waveform-approximant parameter $\pi(\omega)$. For our uninformative choice above, $\pi_{A/B}=1$. But, if cogent prior information about the models is available, this can be included in the prior on $\omega$.

\bibliography{bibliography}

\clearpage

\section{Supplementary Material}
\label{sec:supplementary}

\subsection{Interpreting the Odds}
\changedtwo{In this work, we discuss the odds between different waveform models of binary neutron star collisions. We find that the odds are in favour (i.e.\ greater than unity) of the \teo waveform model across multiple events. The consistency of this finding means that when we combine the odds, we end up with an odds ranging from 1.7 to 2.3 with an uncertainty of $\sim 0.2$. In this Appendix, we discuss the interpretation of this using a simple toy model for demonstration.

\begin{figure}
    \centering
    \includegraphics[width=0.45\textwidth]{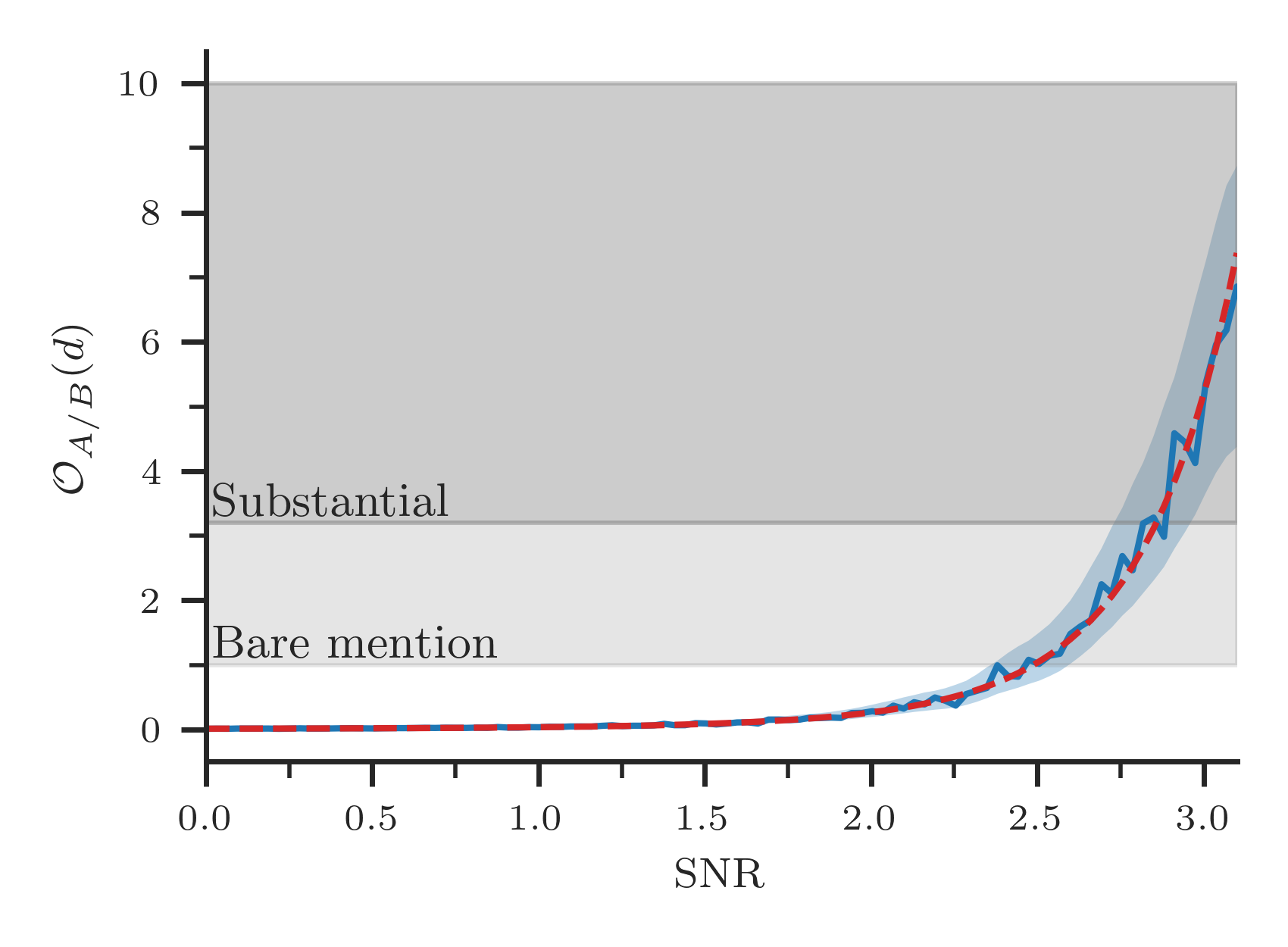}
    \caption{\changedtwo{The odds as a function of the SNR ($\mu/\sigma$) for our toy model as predicted by Eq.~\ref{eqn:odds} (red dashed lines) and by Nested Sampling (blue solid line) and the 3 standard-deviation uncertainty on the Nested Sampling odds (light blue shaded band). In gray bands, we show common thresholds for interpretation of the Odds\cite{Kass:1995loi}.}
    \label{fig:toymodel}}
\end{figure}

The Odds\footnote{Here, we discuss the interpretation of the Odds, but we note that the references refer instead to the Bayes factor. In our case, where the prior-odds are uninformative, these are equivalent. But, one can apply the same interpretation to the odds in the general case that the prior odds are informative.}, quantify the ratio of probabilities of two models given data (see, e.g.~Eq.~\ref{eqn:odds}).
Often, the reader will want to interpret this quantitative statement.
To this end, interpretative thresholds have been defined. One popular approach\cite{Kass:1995loi}, which builds on Jeffrey's foundational work\cite{jeffreys1998theory}, suggests that an odds in the range of 1-3.2 is ``not worth more than a bare mention'' while 3.2-10 is ``substantial evidence``. 
Such a categorisation is useful in ensuring consistency in standards of evidence in scientific investigations.
However, there are several caveats to keep in mind.
First, as noted by the authors, the interpretations should depend on the context. E.g., one may desire a more stringent interpretation for forensic evidence in a criminal trial than other contexts.
Second, the uncertainty on the evidence plays an important role.
If the odds are consistent with unity (to within the stated uncertainty), then no preference between the models can be inferred: we must be able to confidently rule out an odds of unity to have any meaningful result.
Finally, applying blindly the (somewhat arbitrary) thresholds of any particular interpretative table disregards that the odds are (typically) smooth functions of the \emph{strength} of the evidence.

To put these points in context, we consider a simple toy model consisting of data $d=\mu + \epsilon$ where $\epsilon$ is the \emph{noise} drawn from a zero-mean normal distribution with variance $\sigma^2$ and $\mu$ is the mean.
Then, we define two models, model A, where $\mu$ is an unknown parameter to be marginalized over, and model B where $\mu=0$.
Given observed data $d$, the odds of model A vs B (assuming equal prior-odds) are
\begin{equation}
    \odds_{A/B}(d) \approx \frac{\sigma \sqrt{2\pi}}{\mu_{\rm max} - \mu_{\rm min}}e^{d^2 / 2\sigma^2}\,
    \label{eqn:toymodel}
\end{equation}
where $[\mu_{\rm min}, \mu_{\rm max}]$ is the prior range and assumed to be wide with respect to the posterior distribution on $\mu$.
Defining $\sigma/\mu$ as the signal-to-noise ratio (SNR), then in the limit SNR$\gg 1$, we recover the familiar result $\ln \odds_{A/B} \sim \textrm{SNR}^2$.
To simulate our practical settings, in which we cannot calculate the odds in closed form, we estimate the odds using Nested Sampling and plot the results along with the prediction of Eq.~\ref{eqn:toymodel} in Fig.~\ref{fig:toymodel}.
The use of Nested Sampling to approximate the evidence naturally yields a quantified uncertainty on the odds which we plot as a shaded band.

Fig.~\ref{fig:toymodel} demonstrates the final two caveats that we noted above: the uncertainty inherent in estimated odds blurs the sharp boundaries and, regardless of any threshold-based interpretive table, the odds are still a smooth function of the strength of the evidence.
The importance of this for the present work is exemplified by the distinct region in Fig.~\ref{fig:toymodel} in which the odds are distinct from unity (in the sense that the lower bound on the odds is larger than 1), but do not rise to the level of ``substantial''.
How then should be interpret such a result?
Clearly this is not ``substantial'' evidence, but discarding the result seems to ignore that we can confidently exclude an odds of unity.
In this work, we choose to interpret this as ``marginal evidence''.

Finally, we come back to the first caveat, the context. With the next observing run of the international network of gravitational wave detectors expected in the coming years, it is likely that we will see several binary neutron star mergers. 
If just two or three of these are as loud as \seventeen, they may provide sufficient enough evidence for the joint odds to reach the level of substantial evidence. 
However, in the meantime, we anticipate that waveform modellers will improve their model in anticipation of the new observations.
As such, the marginal evidence reported here may be of critical use in determining which aspects of their models to improve!}

\subsection{Posterior distributions for \nineteen and \subthres}
\label{sec:posterior}
\begin{figure*}[h]
    \centering
    \includegraphics[]{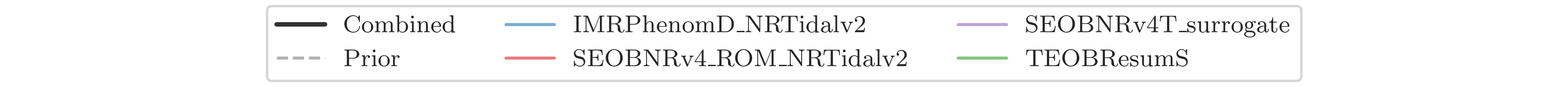}
    \includegraphics[width=0.33\textwidth]{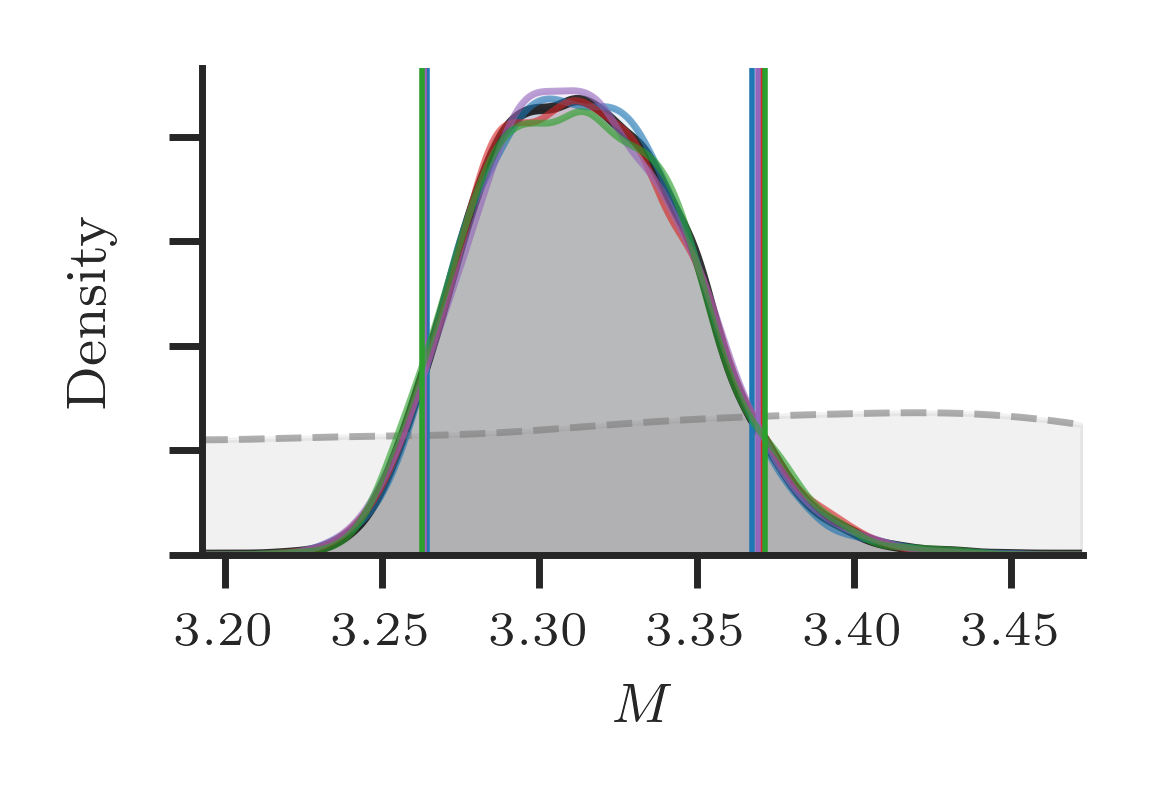}
    \includegraphics[width=0.33\textwidth]{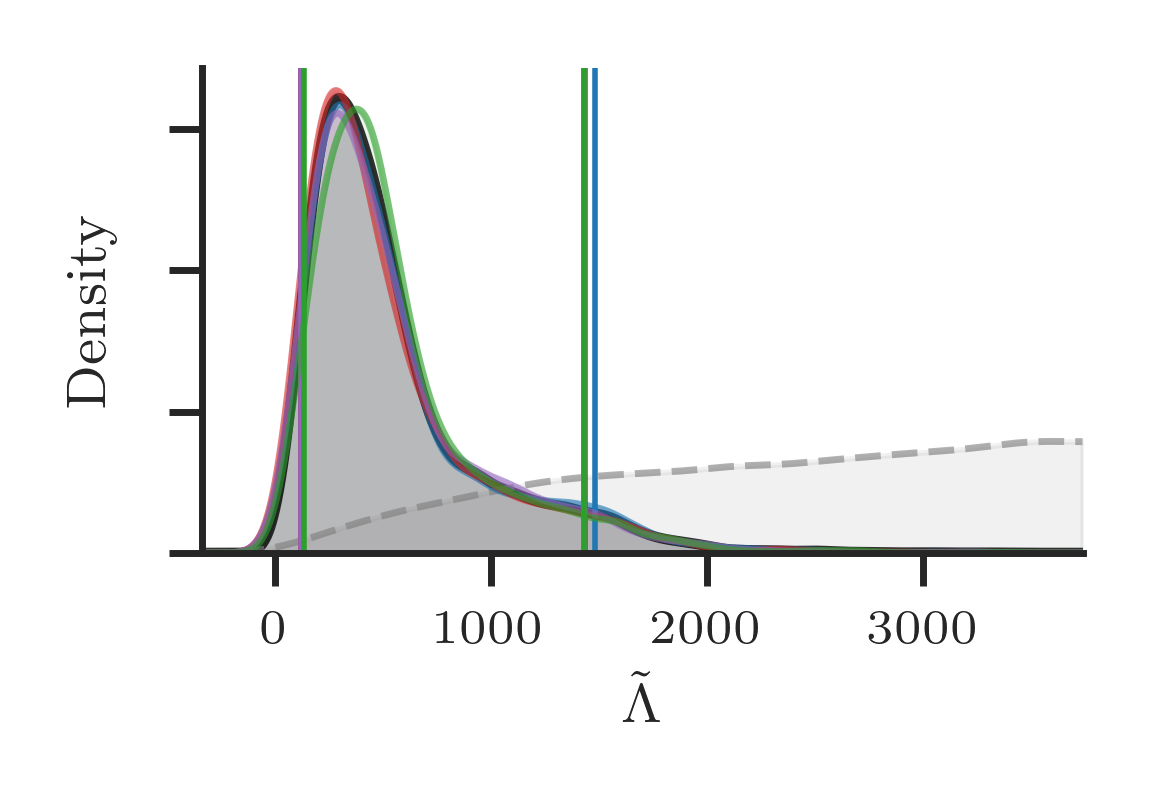}
    \includegraphics[width=0.33\textwidth]{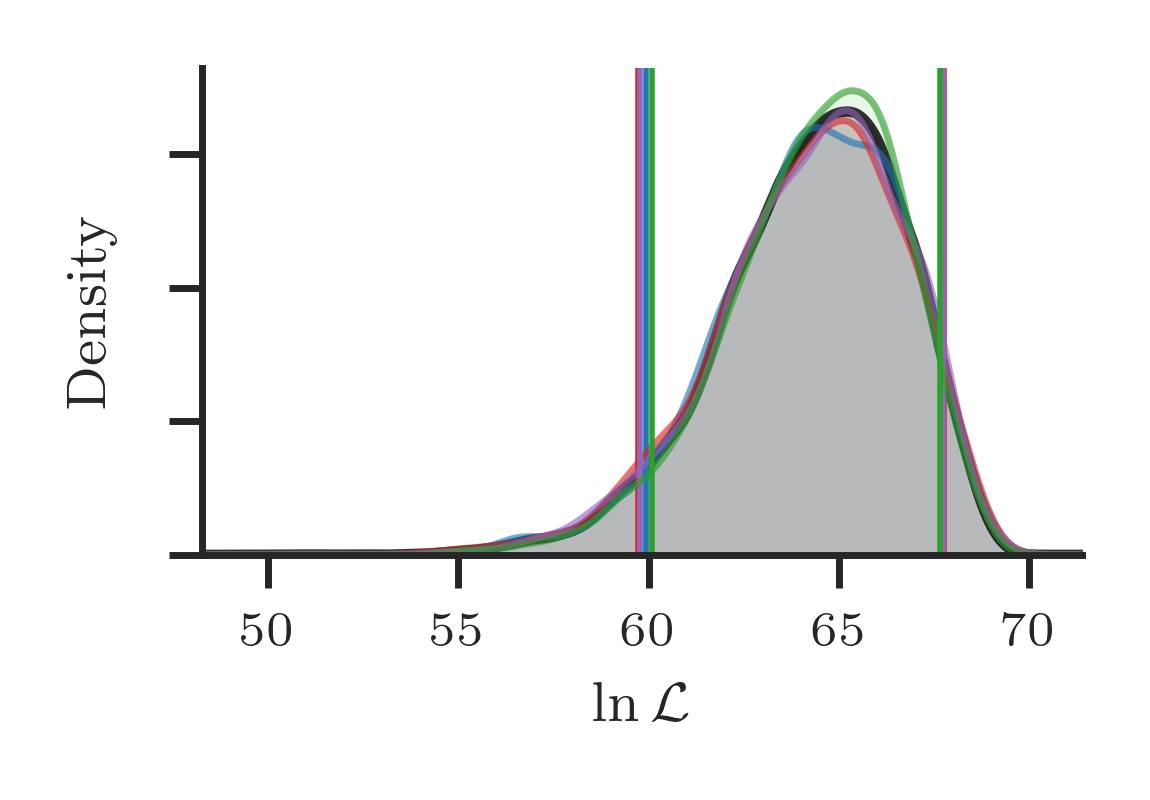}
    \includegraphics[width=0.33\textwidth]{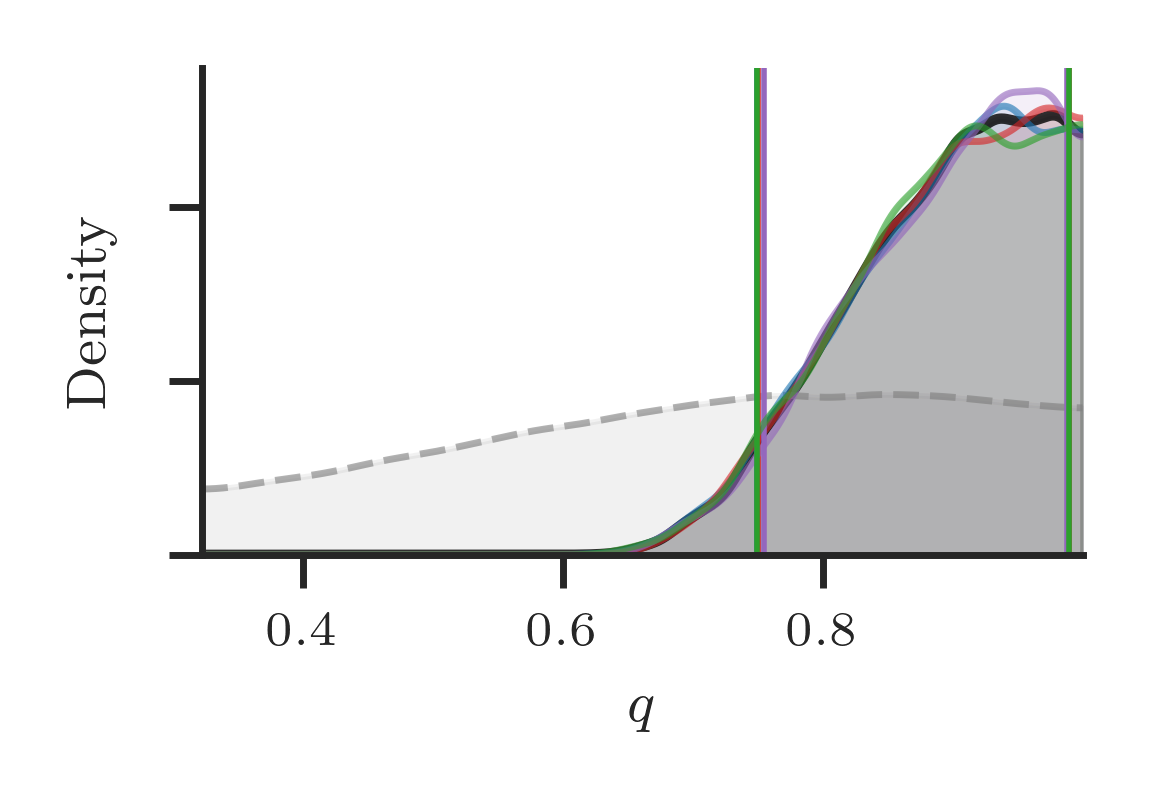}
    \includegraphics[width=0.33\textwidth]{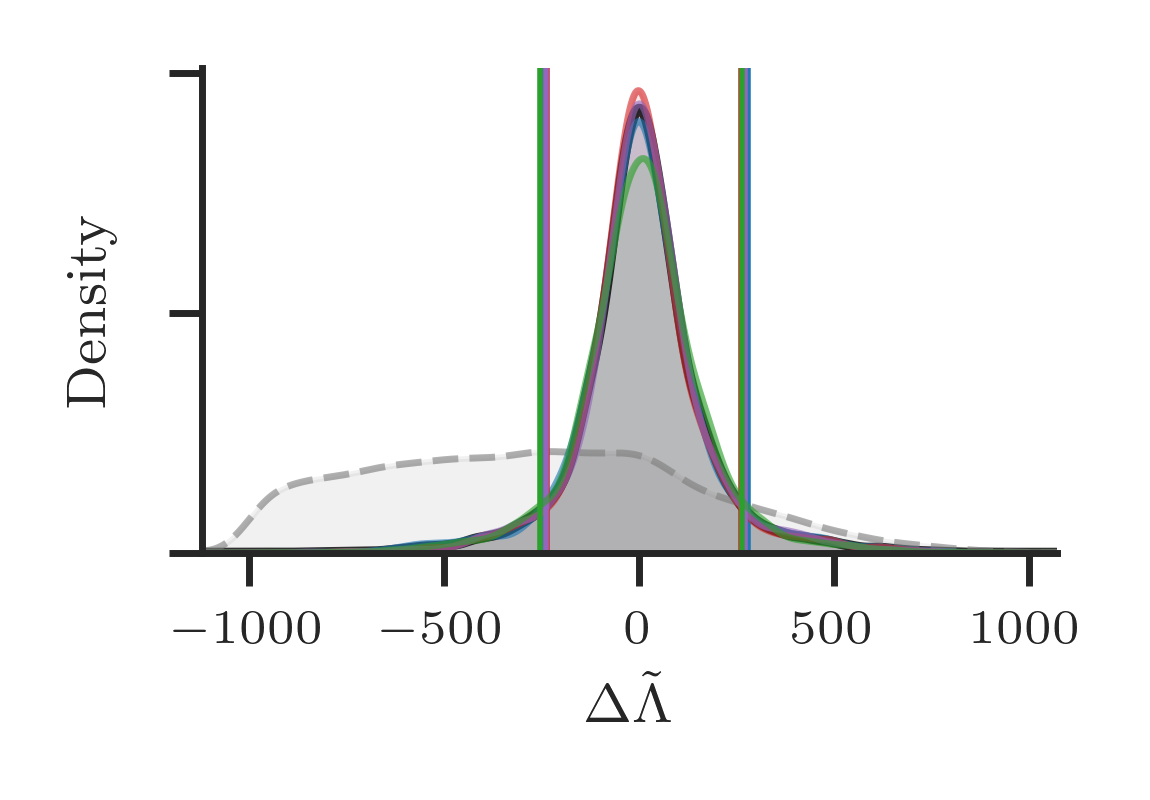}
    \includegraphics[width=0.33\textwidth]{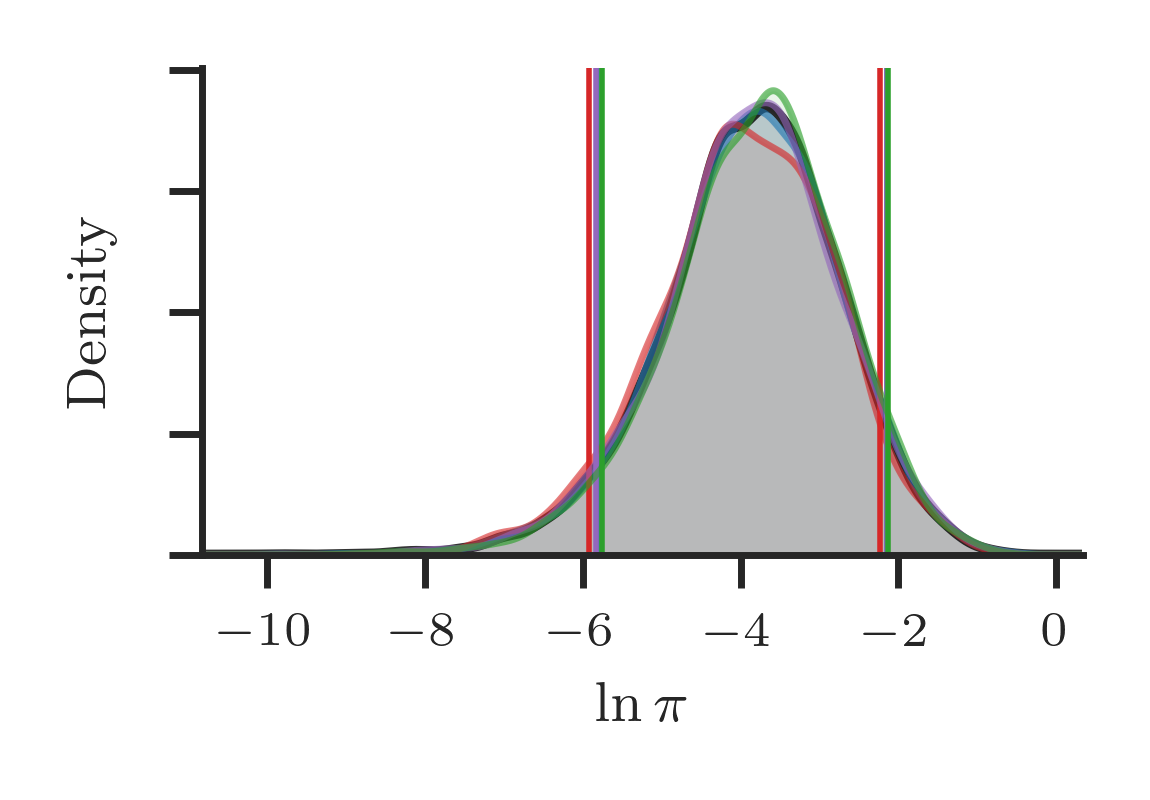}
    \caption{
    Posterior densities for the mass and  tidal deformability of \nineteen. We show the combined posteriors, marginalized over all waveform models using the hypermodel approach, and the posteriors for each individual model, extracted from the combined posterior, as well.
    The dashed curve provides the prior distribution estimated by drawing samples.
    In the right-hand column, we include the distributions of the log-likelihood and log-prior of the posterior samples.}
    \label{fig:GW190425}
\end{figure*}

\begin{figure*}[t]
    \centering
    \includegraphics[]{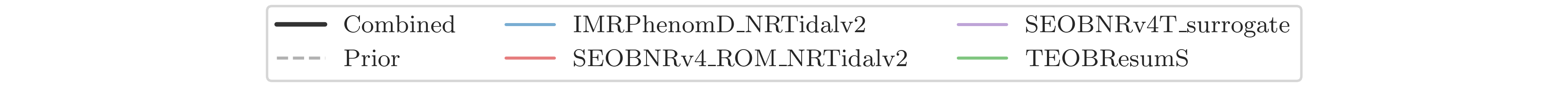}
    \includegraphics[width=0.33\textwidth]{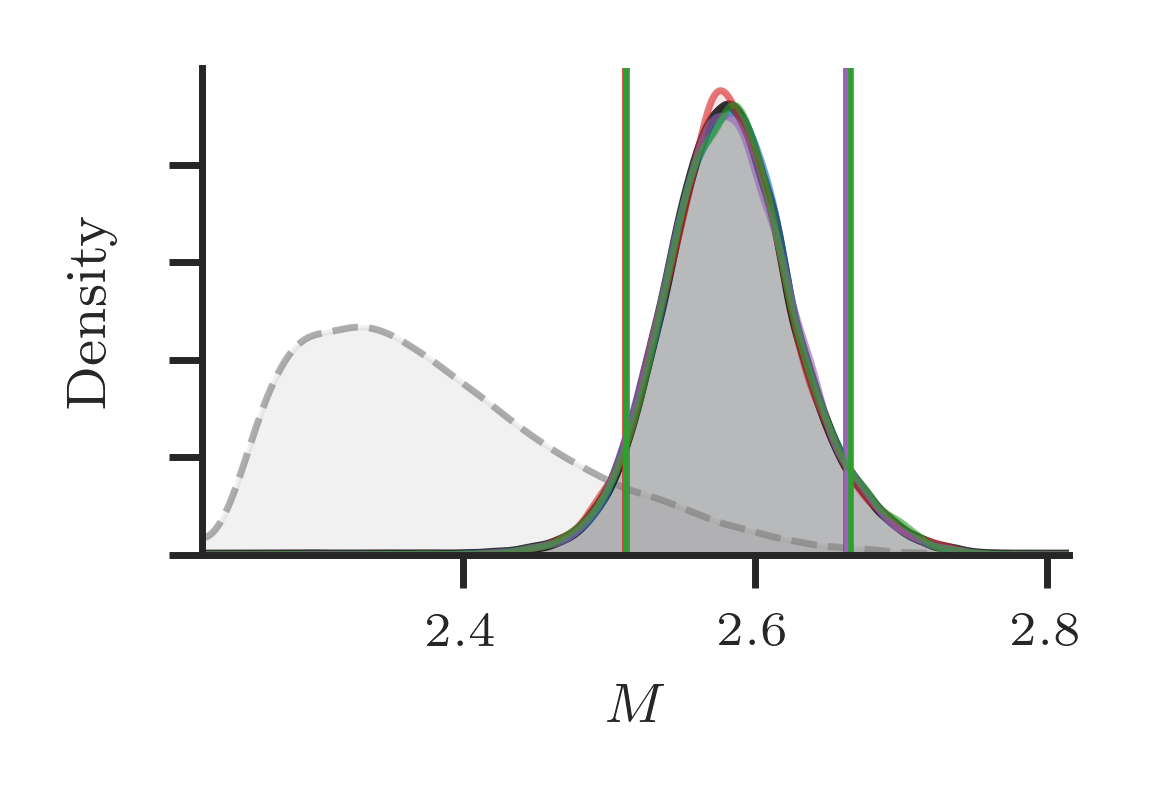}
    \includegraphics[width=0.33\textwidth]{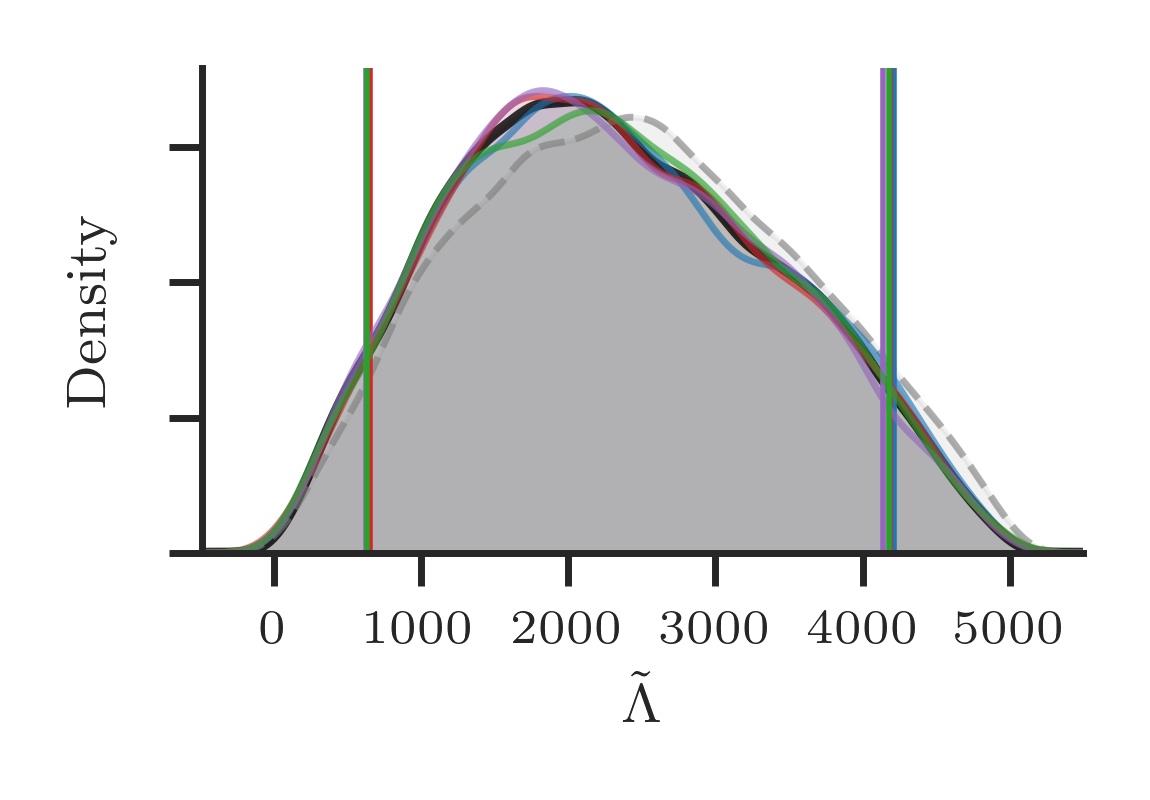}
    \includegraphics[width=0.33\textwidth]{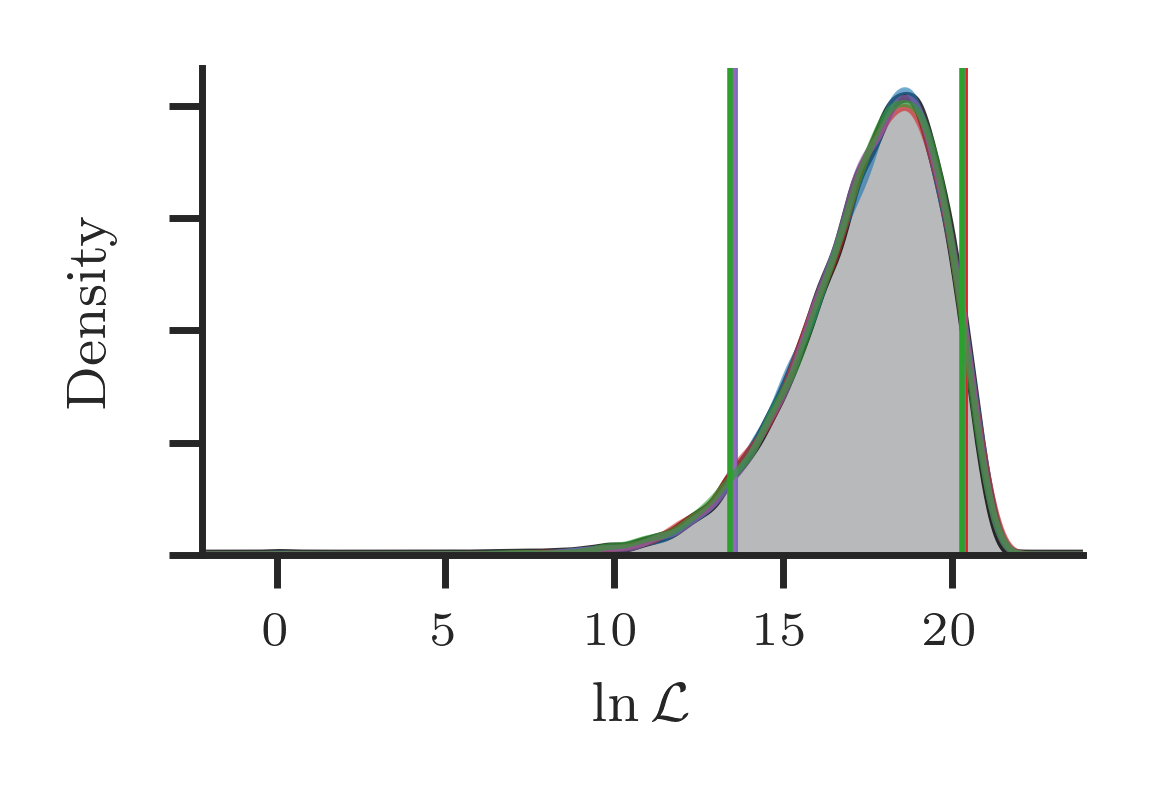}
    \includegraphics[width=0.33\textwidth]{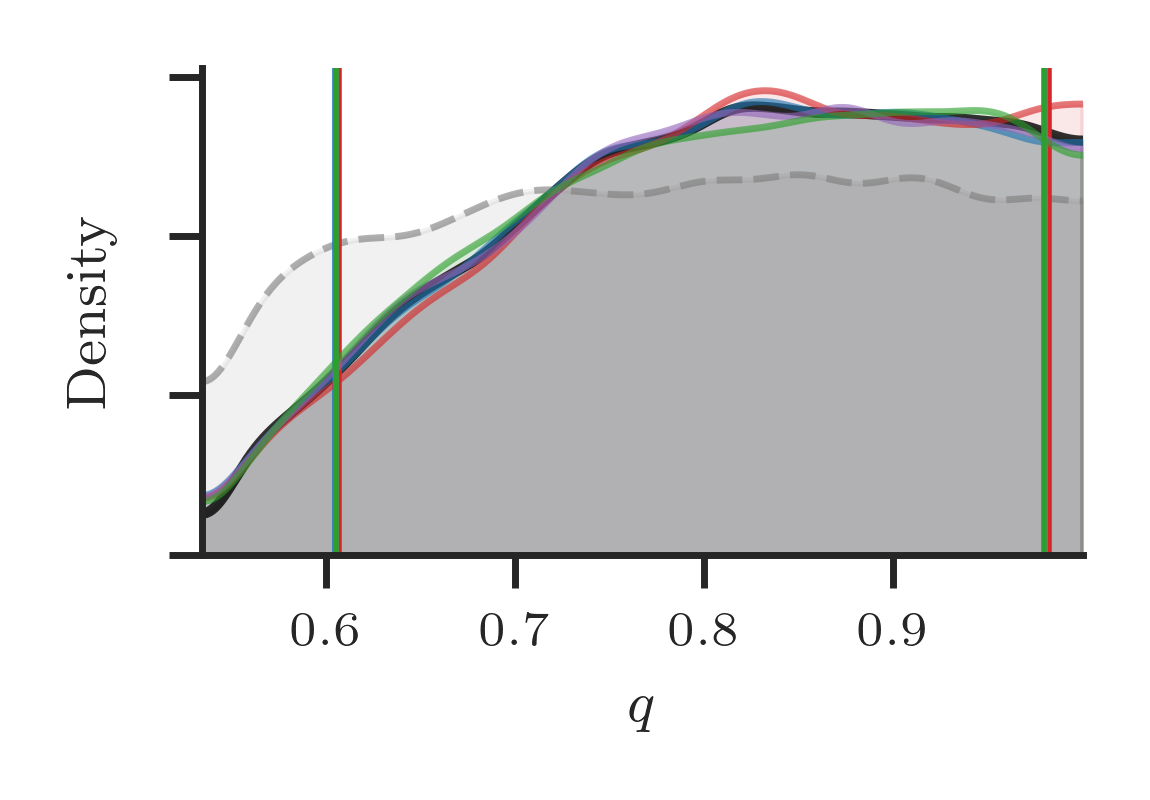}
    \includegraphics[width=0.33\textwidth]{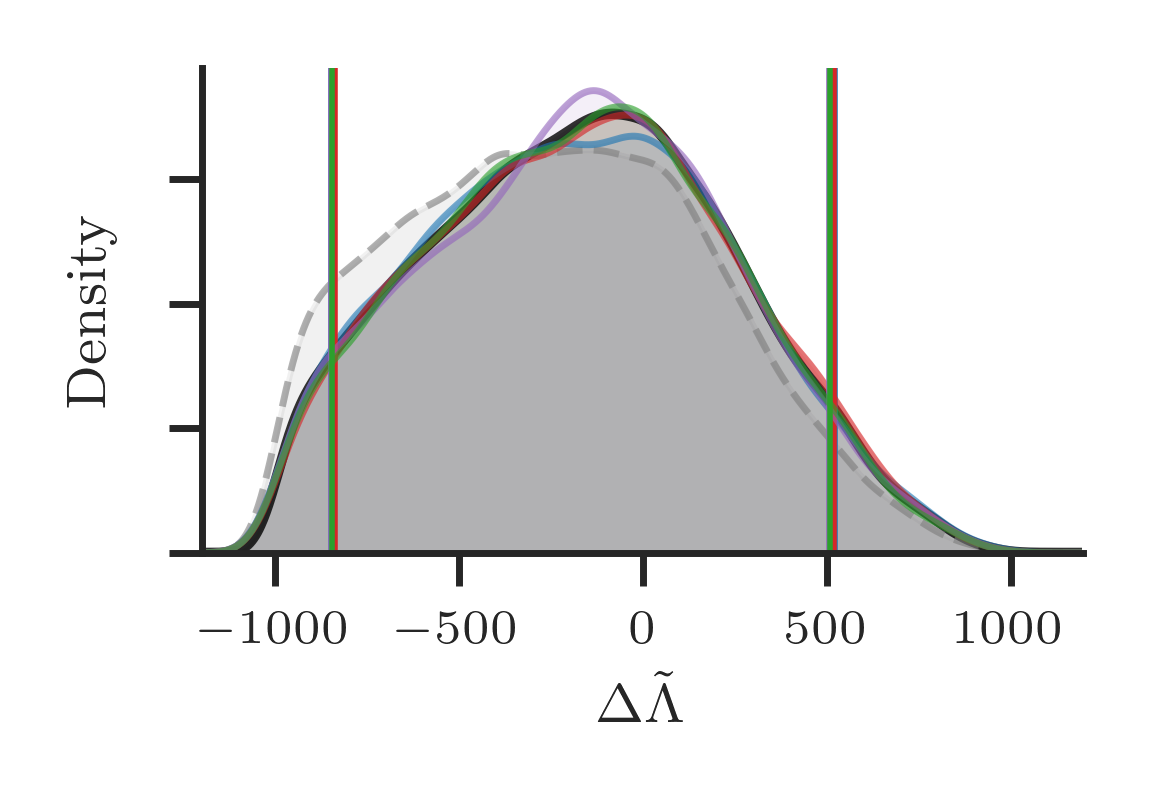}
    \includegraphics[width=0.33\textwidth]{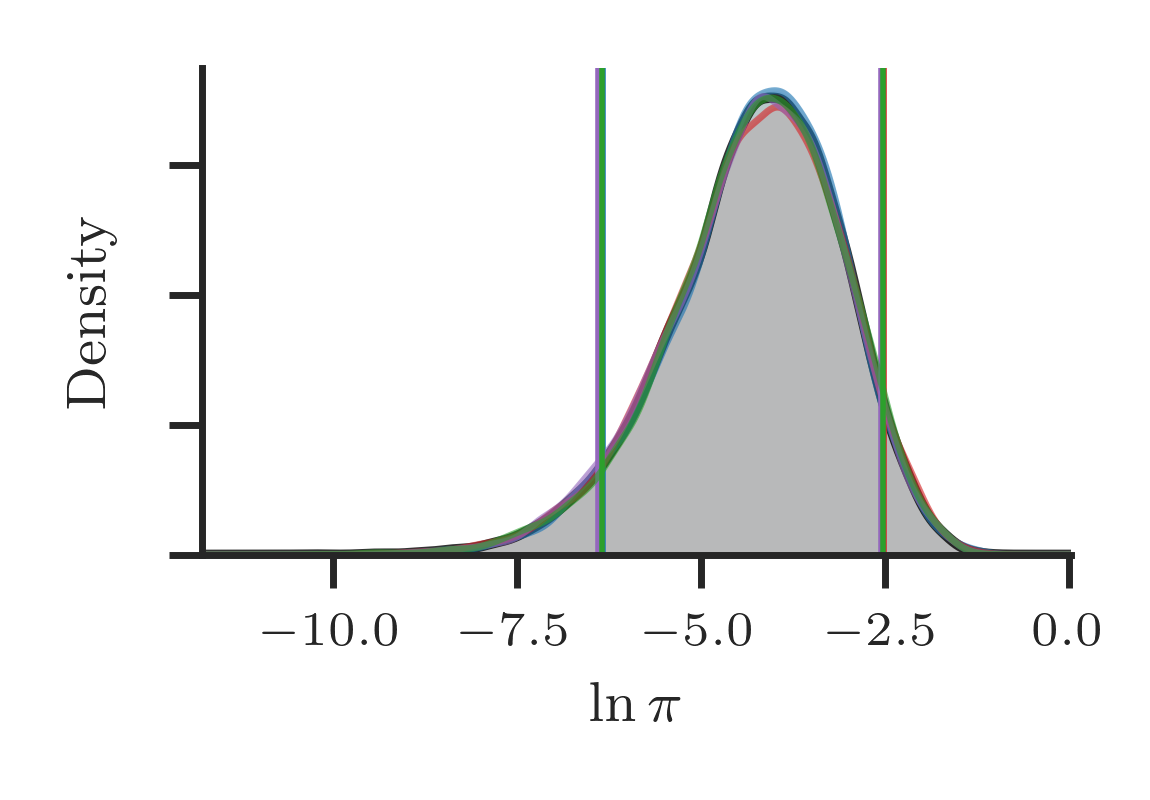}
    \caption{
    Posterior densities for the mass and  tidal deformability of \subthres. We show the combined posteriors, marginalized over all waveform models using the hypermodel approach, and the posteriors for each individual model, extracted from the combined posterior, as well.
    The dashed curve provides the prior distribution estimated by drawing samples.
    In the right-hand column, we include the distributions of the log-likelihood and log-prior of the posterior samples.}
    \label{fig:S200311ba}
\end{figure*}

\changedtwo{Posterior distributions for \nineteen and \subthres are presented in Fig.~\ref{fig:GW190425} and Fig.~\ref{fig:S200311ba}, respectively. }

\section{Data Availability} 

The results for the primary analyses of \seventeen, \nineteen, and \subthres are available in the data release\cite{ashton_gregory_2021_data_release}.

\section{Code Availability} 

The program for the primary analyses of \seventeen, \nineteen, and \subthres is described and available in the data release\cite{ashton_gregory_2021_data_release}.

\section{Acknowledgements}
We thank Sarp Akcay for valuable comments on the manuscript
and comments about the tidal sector of the \teo model.
We are also grateful for discussions with Sebastiano Bernuzzi, Rosella Gamba, and Alessandro Nagar. 
Finally, we thank Jacopo Tissino for pointing out a mistake in Equation~\ref{eqn:sigma} in an early draft of this work.
All Nested Sampling analyses made use of the \texttt{dynesty} package\cite{Speagle:2019ivv} and the higher-order mode analysis of \teo additionally used the massively-parallelised software \texttt{parallel\_bilby}\cite{Smith:2019ucc}.
GA thanks the UKRI Future Leaders Fellowship for support through the grant MR/T01881X/1. TD thanks the Max Planck Society for financial support.
We are grateful for computational resources provided by Cardiff University, and funded by an STFC grant ST/I006285/1 supporting UK Involvement in the Operation of Advanced LIGO.
This work makes use of the
\texttt{scipy}\cite{scipy:2020} and
\texttt{numpy}\cite{oliphant2006guide, van2011numpy, harris2020array}
packages for data analysis and visualisation.

\section{Author contributions} 
Conceptualisation: GA, TD;
Methodology: GA, TD;
Data curation: GA;
Software: GA;
Validation: GA, TD;
Formal analysis: GA;
Resources: GA, TD;
Funding acquisition: GA, TD;
Project administration: GA, TD;
Supervision: GA, TD;
Visualisation: GA;
Writing--original draft: GA, TD;
Writing--review and editing: GA, TD.

\section{Competing interests}  The authors declare that they have no
competing financial interests.

\section{Correspondence} gregory.ashton@ligo.org

\end{document}

%% file: macros.tex
\newcommand{\seventeenImrphenomdNrtidalvtwoPercentage}{23.2}
\newcommand{\seventeenImrphenomdNrtidalvtwoPercentageUncertainty}{0.6}
\newcommand{\seventeenMcmcOddsTeobresumsVsImrphenomdNrtidalvtwo}{1.40}
\newcommand{\seventeenMcmcOddsUncertaintyTeobresumsVsImrphenomdNrtidalvtwo}{0.04}
\newcommand{\seventeenSeobnrvfourRomNrtidalvtwoPercentage}{20.5}
\newcommand{\seventeenSeobnrvfourRomNrtidalvtwoPercentageUncertainty}{0.6}
\newcommand{\seventeenMcmcOddsTeobresumsVsSeobnrvfourRomNrtidalvtwo}{1.59}
\newcommand{\seventeenMcmcOddsUncertaintyTeobresumsVsSeobnrvfourRomNrtidalvtwo}{0.05}
\newcommand{\seventeenSeobnrvfourtSurrogatePercentage}{23.8}
\newcommand{\seventeenSeobnrvfourtSurrogatePercentageUncertainty}{0.6}
\newcommand{\seventeenMcmcOddsTeobresumsVsSeobnrvfourtSurrogate}{1.37}
\newcommand{\seventeenMcmcOddsUncertaintyTeobresumsVsSeobnrvfourtSurrogate}{0.04}
\newcommand{\seventeenTeobresumsPercentage}{32.5}
\newcommand{\seventeenTeobresumsPercentageUncertainty}{0.6}
\newcommand{\seventeenMcmcOddsTeobresumsVsTeobresums}{1.00}
\newcommand{\seventeenMcmcOddsUncertaintyTeobresumsVsTeobresums}{0.02}
\newcommand{\nineteenImrphenomdNrtidalvtwoPercentage}{25.0}
\newcommand{\nineteenImrphenomdNrtidalvtwoPercentageUncertainty}{0.7}
\newcommand{\nineteenMcmcOddsTeobresumsVsImrphenomdNrtidalvtwo}{1.20}
\newcommand{\nineteenMcmcOddsUncertaintyTeobresumsVsImrphenomdNrtidalvtwo}{0.04}
\newcommand{\nineteenSeobnrvfourRomNrtidalvtwoPercentage}{21.3}
\newcommand{\nineteenSeobnrvfourRomNrtidalvtwoPercentageUncertainty}{0.7}
\newcommand{\nineteenMcmcOddsTeobresumsVsSeobnrvfourRomNrtidalvtwo}{1.41}
\newcommand{\nineteenMcmcOddsUncertaintyTeobresumsVsSeobnrvfourRomNrtidalvtwo}{0.06}
\newcommand{\nineteenSeobnrvfourtSurrogatePercentage}{23.5}
\newcommand{\nineteenSeobnrvfourtSurrogatePercentageUncertainty}{0.7}
\newcommand{\nineteenMcmcOddsTeobresumsVsSeobnrvfourtSurrogate}{1.28}
\newcommand{\nineteenMcmcOddsUncertaintyTeobresumsVsSeobnrvfourtSurrogate}{0.05}
\newcommand{\nineteenTeobresumsPercentage}{30.1}
\newcommand{\nineteenTeobresumsPercentageUncertainty}{0.7}
\newcommand{\nineteenMcmcOddsTeobresumsVsTeobresums}{1.00}
\newcommand{\nineteenMcmcOddsUncertaintyTeobresumsVsTeobresums}{0.03}
\newcommand{\subthresImrphenomdNrtidalvtwoPercentage}{25.3}
\newcommand{\subthresImrphenomdNrtidalvtwoPercentageUncertainty}{0.4}
\newcommand{\subthresMcmcOddsTeobresumsVsImrphenomdNrtidalvtwo}{0.99}
\newcommand{\subthresMcmcOddsUncertaintyTeobresumsVsImrphenomdNrtidalvtwo}{0.02}
\newcommand{\subthresSeobnrvfourRomNrtidalvtwoPercentage}{25.0}
\newcommand{\subthresSeobnrvfourRomNrtidalvtwoPercentageUncertainty}{0.4}
\newcommand{\subthresMcmcOddsTeobresumsVsSeobnrvfourRomNrtidalvtwo}{1.01}
\newcommand{\subthresMcmcOddsUncertaintyTeobresumsVsSeobnrvfourRomNrtidalvtwo}{0.02}
\newcommand{\subthresSeobnrvfourtSurrogatePercentage}{24.5}
\newcommand{\subthresSeobnrvfourtSurrogatePercentageUncertainty}{0.4}
\newcommand{\subthresMcmcOddsTeobresumsVsSeobnrvfourtSurrogate}{1.03}
\newcommand{\subthresMcmcOddsUncertaintyTeobresumsVsSeobnrvfourtSurrogate}{0.02}
\newcommand{\subthresTeobresumsPercentage}{25.2}
\newcommand{\subthresTeobresumsPercentageUncertainty}{0.4}
\newcommand{\subthresMcmcOddsTeobresumsVsTeobresums}{1.00}
\newcommand{\subthresMcmcOddsUncertaintyTeobresumsVsTeobresums}{0.02}
\newcommand{\seventeenDynestyOddsTeobresumsVsImrphenomd}{1.6}
\newcommand{\seventeenDynestyOddsUncertaintyTeobresumsVsImrphenomd}{0.3}
\newcommand{\seventeenDynestyOddsTeobresumsVsTeobresums}{1.0}
\newcommand{\seventeenDynestyOddsUncertaintyTeobresumsVsTeobresums}{0.2}
\newcommand{\seventeenDynestyOddsTeobresumsVsSeobnrvfour}{1.9}
\newcommand{\seventeenDynestyOddsUncertaintyTeobresumsVsSeobnrvfour}{0.4}
\newcommand{\seventeenDynestyOddsTeobresumsVsSeobnrvfourt}{1.4}
\newcommand{\seventeenDynestyOddsUncertaintyTeobresumsVsSeobnrvfourt}{0.3}